\documentclass[journal=ancac3,manuscript=article]{achemso}
\usepackage{siunitx}
\DeclareSIUnit{\molar}{M}
\DeclareSIUnit{\Gauss}{G}
\usepackage{amsmath}
\usepackage{amssymb}
\usepackage{placeins}
\usepackage{graphicx}
\usepackage[labelfont=bf]{caption}
\DeclareCaptionFont{xbf}{\bfseries\boldmath}
\usepackage{tabularx}
\usepackage{hyperref}
\usepackage{indentfirst}
\usepackage[utf8]{inputenc}
\usepackage{xcolor}
\usepackage[version=3]{mhchem} 

\author{Fabian A. Freire-Moschovitis}
\affiliation[TUM]
{Technical University of Munich, TUM School of Natural Sciences, Department of
Chemistry, Lichtenbergstra{\ss}e 4, 85748 Garching, Germany}
\alsoaffiliation[MCQST]
{Munich Center for Quantum Science and Technology (MCQST), Schellingstra{\ss}e 4, 80799 M\"unchen, Germany}
\author{Roberto Rizzato}
\affiliation[TUM]
{Technical University of Munich, TUM School of Natural Sciences, Department of
Chemistry, Lichtenbergstra{\ss}e 4, 85748 Garching, Germany}
\alsoaffiliation[MCQST]
{Munich Center for Quantum Science and Technology (MCQST), Schellingstra{\ss}e 4, 80799 M\"unchen, Germany}
\author{Anton Pershin}
\affiliation[Budapest]
{Wigner Research Centre for Physics, Institute for Solid State Physics and Optics, PO. Box 49, Budapest H-1525, Hungary}
\author{Moritz R. Schepp}
\affiliation[TUM]
{Technical University of Munich, TUM School of Natural Sciences, Department of
Chemistry, Lichtenbergstra{\ss}e 4, 85748 Garching, Germany}
\author{Robin D. Allert}
\affiliation[TUM]
{Technical University of Munich, TUM School of Natural Sciences, Department of
Chemistry, Lichtenbergstra{\ss}e 4, 85748 Garching, Germany}
\alsoaffiliation[MCQST]
{Munich Center for Quantum Science and Technology (MCQST), Schellingstra{\ss}e 4, 80799 M\"unchen, Germany}
\author{Lina M. Todenhagen}
\affiliation[WSI]
{Technical University of Munich, Walter Schottky Institut, Am Coulombwall 4, 85748 Garching, Germany}
\alsoaffiliation[Physics]
{Technical University of Munich, TUM School of Natural Sciences, Department of
Physics, James-Franck-Stra{\ss}e 1, 85748 Garching, Germany}
\alsoaffiliation[MCQST]
{Munich Center for Quantum Science and Technology (MCQST), Schellingstra{\ss}e 4, 80799 M\"unchen, Germany}
\author{Martin S. Brandt}
\affiliation[WSI]
{Technical University of Munich, Walter Schottky Institut, Am Coulombwall 4, 85748 Garching, Germany}
\alsoaffiliation[Physics]
{Technical University of Munich, TUM School of Natural Sciences, Department of
Physics, James-Franck-Stra{\ss}e 1, 85748 Garching, Germany}
\alsoaffiliation[MCQST]
{Munich Center for Quantum Science and Technology (MCQST), Schellingstra{\ss}e 4, 80799 M\"unchen, Germany}
\author{Adam Gali}
\affiliation[Budapest]
{Wigner Research Centre for Physics, Institute for Solid State Physics and Optics, PO. Box 49, Budapest H-1525, Hungary}
\alsoaffiliation
{Department of Atomic Physics, Institute of Physics, Budapest University of Technology and Economics, M\H{u}egyetem rakpart 3, Budapest H-1111, Hungary}
\author{Dominik B. Bucher}
\affiliation[TUM]
{Technical University of Munich, TUM School of Natural Sciences, Department of
Chemistry, Lichtenbergstra{\ss}e 4, 85748 Garching, Germany}
\alsoaffiliation[MCQST]
{Munich Center for Quantum Science and Technology (MCQST), Schellingstra{\ss}e 4, 80799 M\"unchen, Germany}
\email{dominik.bucher@tum.de}

\title[An \textsf{achemso} demo]{The Role of Electrolytes in the Relaxation of Near-Surface Spin Defects in Diamond}

\begin{document}
\begin{abstract}
Quantum sensing with spin defects in diamond, such as the nitrogen-vacancy (NV) center, enables the detection of various chemical species on the nanoscale. Molecules or ions with unpaired electronic spins are typically probed by their influence on the NV center's spin relaxation. Whereas it is well-known that paramagnetic ions reduce the NV center's relaxation time ($T_1$), here we report on the opposite effect for diamagnetic ions. We demonstrate that millimolar concentrations of aqueous diamagnetic electrolyte solutions increase the $T_1$ time of near-surface NV center ensembles compared to pure water. To elucidate the underlying mechanism of this surprising effect, single and double quantum NV experiments are performed, which indicate a reduction of magnetic and electric noise in the presence of diamagnetic electrolytes. In combination with \textit{ab initio} simulations, we propose that a change in the interfacial band bending due to the formation of an electric double layer leads to a stabilization of fluctuating charges at the interface of an oxidized diamond. This work not only helps to understand noise sources in quantum systems but could also broaden the application space of quantum sensors towards electrolyte sensing in cell biology, neuroscience and electrochemistry.

\end{abstract}

Nitrogen-vacancy (NV) centers in diamond offer a broad platform for quantum sensing applications ranging from the measurement of basic physical properties, such as temperature,\cite{Acosta2010b} pressure,\cite{Ivady2014b} strain,\cite{Teissier2014} electric\cite{Dolde2011b} and magnetic fields\cite{Balasubramanian2008,Maze2008} down to single cells,\cite{Glenn2015a} single molecules,\cite{Lovchinsky2016} or even single nuclear spins.\cite{Muller2014,Sushkov2014b} This high sensitivity is achieved due to the atomic size of the qubit enabling its location only a few nanometer away from the diamond interface (see Figure \ref{fig:Figure1}a).\cite{Bucher2019a} 
Near-surface NV centers ($\leq \SI{20}{\nano\meter}$ below the surface) can be used to sense nuclear magnetic resonance (NMR)\cite{Lovchinsky2016,Lovchinsky2017,Liu2022,Allert2022,Allert2022b} and electron spin resonance (ESR) signals\cite{Shi2015} from nanoscale chemical and biological samples. The NV center translates these magnetic field fluctuations directly to an optical signal, detected by a change in the fluorescence intensity. Together with optical spin state initialization with green laser light and coherent spin state manipulation with microwave pulses, these key features make the NV center a suitable tool for (bio)chemical analysis.\cite{Schirhagl2014a,Bucher2019,Bucher2019a,Allert2022b}

NV centers are also perceptive to electric fields\cite{Dolde2011b,Kim2015a}. Even a single elementary charge located $\sim$ \SI{150}{\nano\meter} away from the quantum sensor produces a strong enough static electric field to affect an NV center's ODMR (optically detected magnetic resonance) transitions.\cite{Dolde2011b} This high sensitivity of NV centers to magnetic or electric fields marks a narrow ridge: On the one hand, it allows for single nuclear spin or elementary charge detection, on the other hand it makes the qubit prone to magnetic or electronic spin noise in its close environment. This is particularly relevant for NV centers close to the surface, where interfacial processes and defects cause additional noise and reduce the performance of the NV-quantum sensors.\cite{Stacey2019,Sangtawesin2019,Romach2015,Chrostoski2018}

Due to the NV center's susceptibility to a broad range of frequencies (from DC to \si{\giga\hertz}),\cite{Degen2017a} noise can be measured by applying different sensing protocols.\cite{Bucher2019} High frequency noise ($\sim$ \si{\giga\hertz}) can be probed by the longitudinal spin-lattice relaxation time ($T_1$) with a protocol that is usually referred to as NV relaxometry.\cite{Bucher2019,Steinert2013a,Schirhagl2014a,Degen2017a,Mzyk2022} The $T_1$ time defines the time constant of the spin-state-dependent fluorescence decay from the magnetic sublevel $m_{s}=0$ (bright state) to thermal equilibrium (mixed state) and is on the order of a few milliseconds for near-surface NV-ensembles in bulk diamonds,\cite{Bucher2019a} or on the order of a few hundred microseconds for nanodiamonds (depending on their size).\cite{Schirhagl2014a} Generally, any magnetic noise overlapping with the NV center's Larmor frequency (on the order of the zero-field splitting $D=\SI{2.87}{\giga\hertz}$) will decrease the $T_1$ relaxation time (see Figure \ref{fig:Figure1}b).\cite{Bucher2019,Steinert2013a,PeronaMartinez2020a,Li2022} Relaxometry has been successfully applied to map high frequency magnetic noise originating from inside the diamond (\textit {\textit {e.g.}}, paramagnetic impurities\cite{Jarmola2012}), at the diamond interface (\textit {e.g.}, dangling bonds\cite{Rosskopf2014a,Stacey2019}) or from samples on top of the diamond (\textit {e.g.}, organic radicals\cite{PeronaMartinez2020a,Nie2022} or paramagnetic ions, such as Mn\textsuperscript{2+},\cite{Ziem2013a} Fe\textsuperscript{3+},\cite{Ermakova2013a} or Gd\textsuperscript{3+}\cite{Steinert2013a,Li2022}). Furthermore, nanoscale NV relaxometry has also been used to determine the pH value\cite{Fujisaku2019} or to monitor chemical reactions \textit{in situ}.\cite{Simpson2017a}

While the increase of the relaxation rate due to paramagnetic species has been studied extensively, herein we detect the opposite effect for diamagnetic ions. When near-surface NV centers in oxidized diamonds are exposed to aqueous diamagnetic electrolytes, we observe a systematic extension of the $T_1$ relaxation time compared to pure (deionized) water. We show that this unexpected effect is proportional to the electrolyte concentration, reversible and dependent on the NV center's implantation depth. In order to shed light on the underlying sensing mechanism, we perform single and double quantum relaxometry experiments which indicate a reduction of electric as well as magnetic noise. Furthermore, double electron electron resonance (DEER) experiments show a similar effect on surface dark spins, which possibly act as surface reporter spins. In combination with theoretical methods including \textit{ab initio} simulations of the diamond/electrolyte interface, we propose that diamagnetic ions alter the interfacial band bending. This leads to a stabilization of fluctuating charges at the interface and to the increase of the $T_1$ relaxation time.  

\section{Results}
\subsection{\boldmath{$T_1$} Relaxometry on Electrolytes with Near-Surface NV center Ensembles}
\begin{figure} [htbp]
    \centering
    \includegraphics{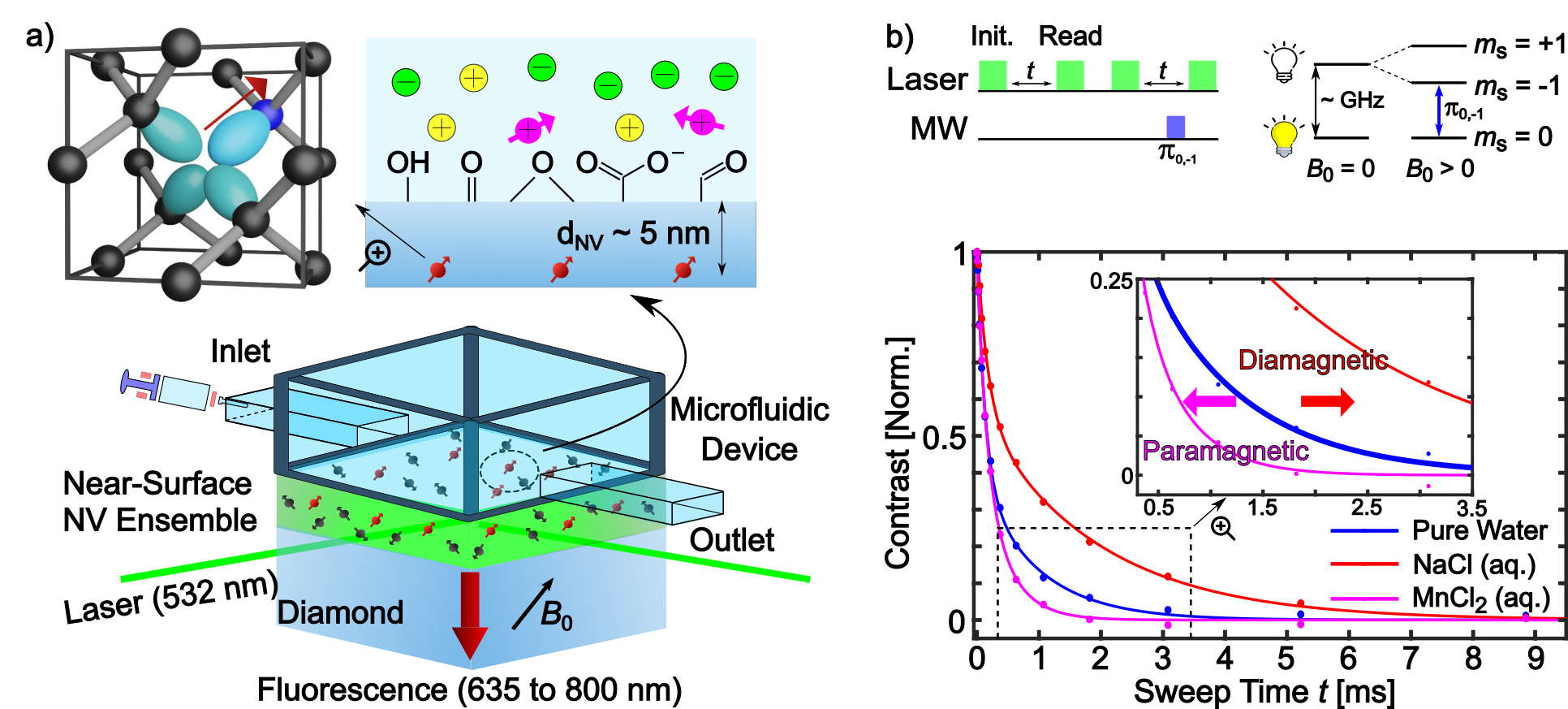}
    \caption{\textbf{Scheme of NV relaxometry experiments with aqueous electrolyte solutions. (a) Top (left): The NV center in the diamond crystal lattice. A nitrogen atom (blue) replaces a carbon atom (black) in the crystal. Together with an adjacent vacancy an NV center is formed. The orbitals (petrol) indicate four possible NV center orientations within the diamond lattice. Top (right): Scheme of an oxidized diamond surface with an ensemble of near-surface NV centers ($\text{d}\textsubscript{NV} \sim \SI[detect-weight]{5}{\nano\meter}$). The surface termination consists of hydroxyl groups, ethers, ketones, aldehydes and carboxylic acids. We probe pure water, paramagnetic (purple) and diamagnetic (yellow) ions. Bottom: A microfluidic device placed on top of the diamond. In- and outlet allow for adding and removing aqueous electrolyte solutions or pure water from the diamond surface. (b) Top (left): NV relaxometry pulse sequence. Two laser pulses for spin state initialization and readout are separated by a sweep time \textit{t}. A consecutive measurement with a $\pi$-pulse at the end is used for noise cancellation.\cite{Bucher2019} Top (right): Energy levels of the NV center's electronic ground state ($S=1$). The zero-field splitting ($D=\SI[detect-weight]{2.87}{\giga\hertz}$) separates the $m_{s}=0$ (bright) and $m_{s}=\pm1$ states (dark). A bias magnetic field $B_0$ splits the degenerate $m_{s}=\pm1$ states according to the Zeeman effect. Bottom: $T_1$ relaxation curves of pure water and solutions of NaCl (\SI[detect-weight]{500}{\milli\molar}) and  MnCl\textsubscript{2} (\SI[detect-weight]{1}{\micro\molar}). While paramagnetic MnCl\textsubscript{2} reduces the $T_1$ time with respect to pure water, diamagnetic NaCl extends the $T_1$ time. Experiments are performed at $f_{\text{NV}}=\SI[detect-weight]{1.88}{\giga\hertz}$.}}
    \label{fig:Figure1}
\end{figure}

In this study we use near-surface high density NV center ensembles (implanted with \textsuperscript{15}N at an energy of 2.5\,keV and a fluence of $2\times10^{12}\, \si{\per\square\centi\meter}$), distributed $\sim \SI{5}{\nano\meter}$ underneath the diamond surface,\cite{Pham2016,JacobHenshaw,Liu2022} to investigate the effect of aqueous electrolyte solutions on the spin-lattice relaxation time $T_1$ probed by NV relaxometry. Before experiments are conducted, we prepare the diamond surface with a triacid clean procedure according to Brown \textit{et al.}\cite{Brown2019} This procedure not only ensures to remove non-diamond carbon material from the interface but also creates an oxidized surface comprised of hydroxyl groups, ethers, ketones, aldehydes and carboxylic acids.\cite{Li2021a} We position the diamond in a microfluidic device that guarantees controllable in- and output of the applied liquids, prevents sample evaporation and provides a constant and defined volume for following measurements (see Figure \ref{fig:Figure1}a).\cite{Allert2022b} Importantly, the microfluidic device avoids a direct contact between the liquid and the microwave delivery.
The NV relaxometry protocol depicted in Figure \ref{fig:Figure1}b essentially consists of two \SI{532}{\nano\meter} laser pulses of \SI{5}{\micro\second} duration for optical spin state initialization and readout separated by a sweep time \textit{t}. A subsequent measurement with a $\pi_\text{0,-1}$-pulse (where the subscripts 0 and -1 indicate transitions between $m_s=0 \leftrightarrow m_s=-1$) at the end is used for normalization and noise cancellation.\cite{Bucher2019} The $T_1$ time can be extracted from the (bi)exponential fit of the relaxation curves depicting the contrast as a function of sweep time \textit{t} (see Supplementary Note 1 for fitting details).

We perform the measurements by filling the microfluidic channel and covering the diamond surface either with pure water or aqueous electrolyte solutions (see Methods for detail). Figure \ref{fig:Figure1}b depicts the $T_1$ relaxation curves when water and solutions of diamagnetic NaCl or paramagnetic MnCl\textsubscript{2} cover the diamond surface. Paramagnetic MnCl\textsubscript{2} (\SI{1}{\micro\molar}) on the diamond leads to a $T_1$ time reduction by a factor of $0.47\pm0.19$ with respect to water, which is in accordance with other studies and can be ascribed to the strong dipole-dipole interaction of the NV center with paramagnetic species.\cite{Ziem2013a,Steinert2013a} In contrast, when we repeat the same experiment with diamagnetic NaCl (\SI{500}{\milli\molar}) solution, we observe an extension of the $T_1$ time by a factor of $2.04\pm0.45$ compared to water. Experiments supporting this observation are also conducted with other diamagnetic salt solutions (mono-, di- and trivalent) and reveal similar results (see Supplementary Note 2). Therefore, we choose NaCl as a representative of a standard diamagnetic electrolyte for the following measurements in our work and expect comparable results for other diamagnetic salt solutions.

Moreover, by tuning the magnetic field $B_0$ and thereby the NV center's Larmor frequency NV\textsubscript{0,-1} (\textit {i.e.}, the $m_s=0 \rightarrow m_s=-1$ transition frequency) we are able to map the spectral noise density. We probe water/NaCl (\SI{500}{\milli\molar}) solution with  NV\textsubscript{0,-1} frequencies from \SI{131}{\mega\hertz} to \SI{2.87}{\giga\hertz} and observe a similar effect over the entire frequency range (see Supplementary Note 2). Consequently, the extension of the $T_1$ time of near-surface NV-ensembles with exposure to diamagnetic electrolyte solutions is an effect that covers a broad range of (high) frequencies (\textit {i.e.}, from $\sim$ hundreds of \si{\mega\hertz} to \si{\giga\hertz}).

Additionally, in order to exclude an impact of the solvent's physical properties (\textit {i.e.}, polarity) on our experiments,\cite{Kim2015a} we choose typical organic solvents whose dielectric constants ($\kappa$) and chemical structure differ significantly from water ($\kappa=80$\cite{Seyferth1987}) and probe them with relaxometry (see Supplementary Note 3). Since the $T_1$ time remains unaffected, we conclude that the herein described effect is not induced by the physical properties of water, but by the diamagnetic electrolyte.
\FloatBarrier

\subsection{Sensitivity of \boldmath{$T_1$} Relaxometry on Electrolytes}
In order to obtain information about the sensitivity of the NV relaxometry protocol to para- and diamagnetic electrolyte solutions, we perform additional measurement series where the electrolyte concentration is increased stepwise (from $10^{-5}$ to $10^{-2}$ \si{\milli\molar} in the case of paramagnetic MnCl\textsubscript{2} and from $10^{-4}$ to $10^{3}$  \si{\milli\molar} in the case of diamagnetic NaCl).

\begin{figure}
    \centering
    \includegraphics{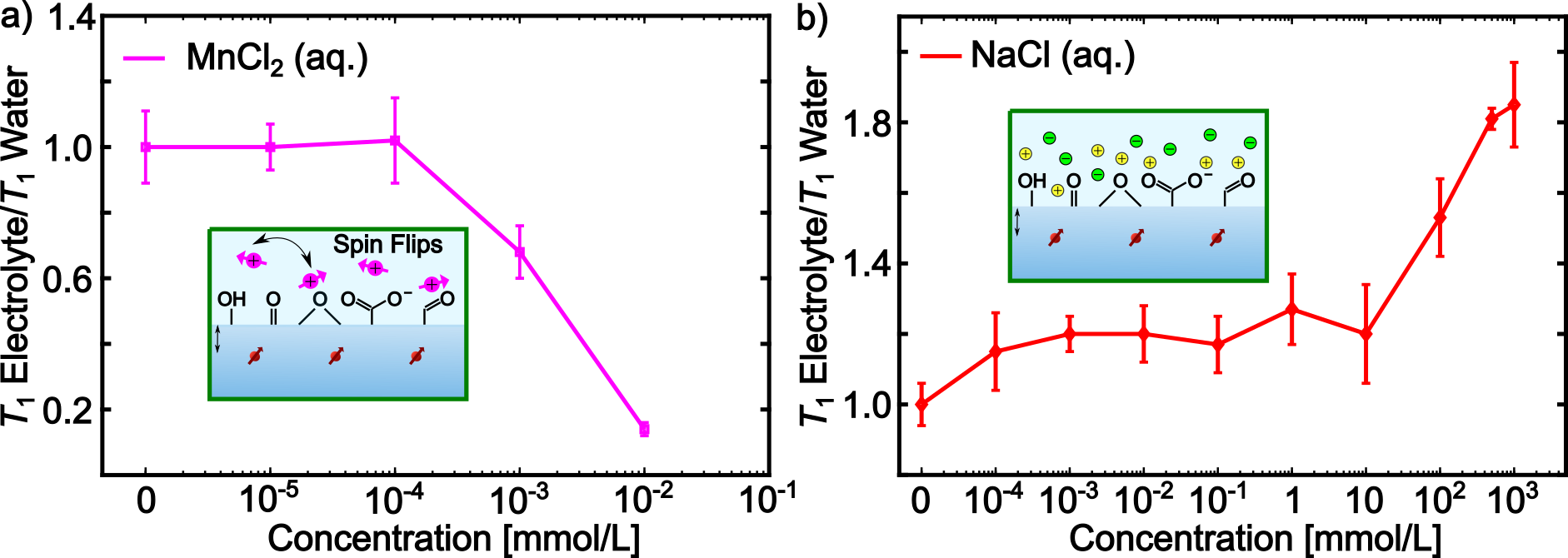}
    \caption{NV relaxometry with increasing concentrations of para- and diamagnetic electrolyte solutions. (a) Paramagnetic  MnCl\textsubscript{2} shows a stepwise $T_1$ time decrease for concentrations in the micromolar regime until the effect reaches a maximum measurable decline of $86\pm10\%$ for \SI[detect-weight]{10}{\micro\molar} solutions with respect to water. (b) In contrast, diamagnetic NaCl (right) shows a slight increase of the $T_1$ time compared to water, which then fluctuates moderately from the micromolar to the lower millimolar regime. For concentrations $\geq \SI[detect-weight]{10}{\milli\molar}$ the $T_1$ time increases gradually along with the NaCl concentration until the effect saturates to $81\pm11\%$ for NaCl (\SI[detect-weight]{500}{\milli\molar}) solutions. Experiments are performed at $f_{\text{NV}}=\SI[detect-weight]{1.88}{\giga\hertz}$.}
    \label{fig:Figure2}
\end{figure}

Paramagnetic MnCl\textsubscript{2} shows a stepwise $T_1$ decrease in micromolar concentrations reaching a decline of up to $86\pm10\%$ for a \SI{10}{\micro\molar} solution with respect to water covering the diamond (see Figure \ref{fig:Figure2}a and Supplementary Note 4). Note that a further concentration increase ($> \SI{10}{\micro\molar}$) is not measurable, as it leads to a collapse of the $T_1$ time. In contrast, diamagnetic NaCl shows a slight $T_1$ increase compared to water, which then fluctuates moderately from micromolar to lower millimolar concentrations. Importantly, a significant and gradual $T_1$ increase is measurable from \SI{10}{\milli\molar} to \SI{500}{\milli\molar} NaCl solution, where the effect saturates at $81\pm11\%$ with respect to water (see Figure \ref{fig:Figure2}b and Supplementary Note 4).

The decrease of the $T_1$ time with paramagnetic species (\textit {e.g.}, MnCl\textsubscript{2}) is expected and well studied.\cite{Steinert2013a,Schirhagl2014a,Rosskopf2014a,Mzyk2022} Here, high frequency ($\sim$ \si{\giga\hertz}) noise originates from magnetic dipole-dipole interactions of the NV center's electronic spin and the sample's electronic spin (``spin-flips''), resulting in a decline of the $T_1$ time if unpaired electrons are near the sensor. However, for diamagnetic ions (\textit {e.g.}, NaCl) these interactions are absent as only paired electrons without a (net) magnetic moment are present. Surprisingly, here we observe a gradual extension of the $T_1$ time with increasing millimolar concentrations of diamagnetic NaCl solution. To test whether this effect is influenced by the valency of the electrolyte, we perform additional relaxometry experiments which indicate a larger impact on the $T_1$ time by (divalent) MgSO\textsubscript{4} compared to (monovalent) NaCl solutions (see Supplementary Note 4).

\FloatBarrier

\subsection{Reversibility, Passivation and NV center Depth}
\begin{figure} [htbp]
    \centering
    \includegraphics{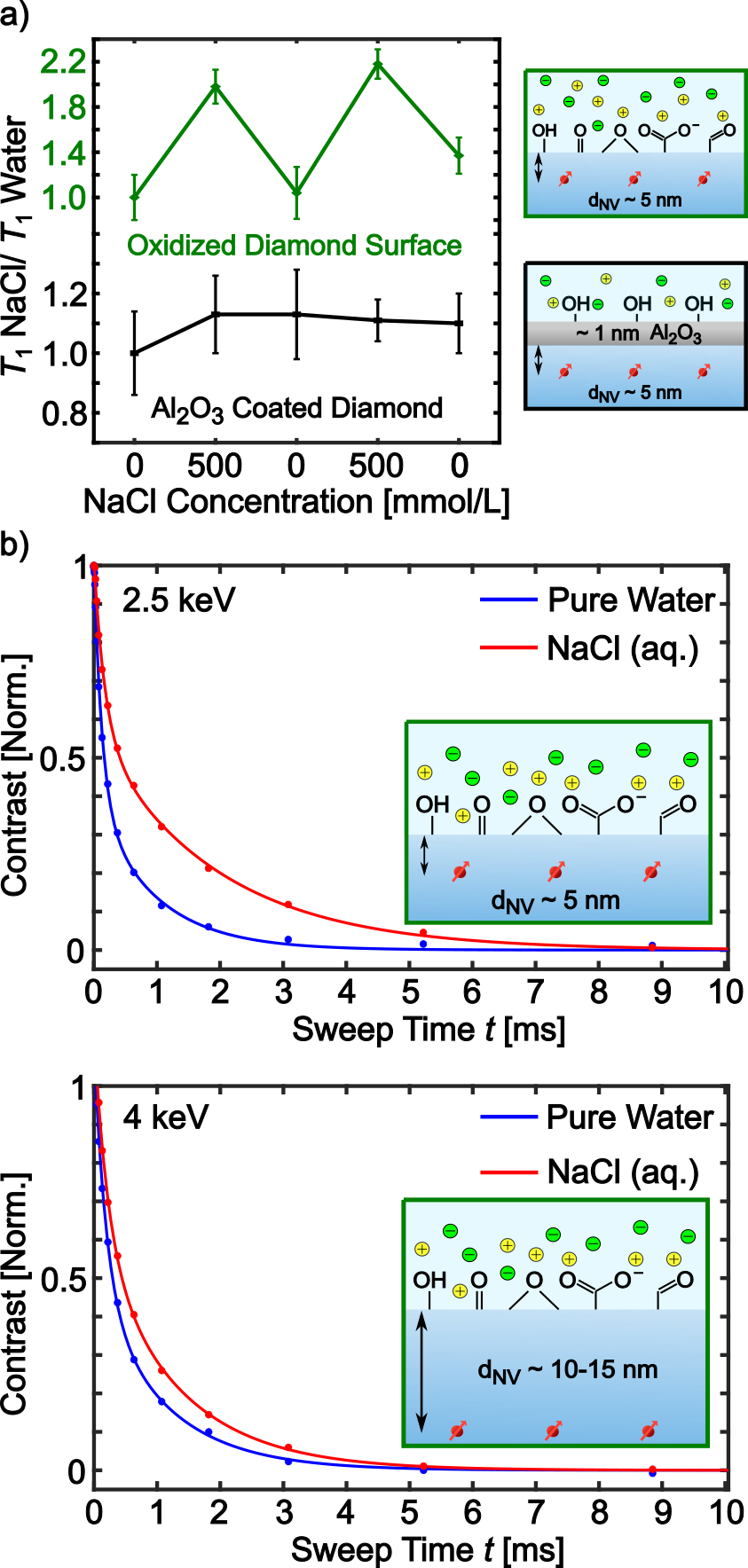}
    \caption{NV relaxometry experiments with different surface terminations and NV center ensemble implantation depths. (a) An oxidized diamond is alternatingly covered with water and NaCl (\SI[detect-weight]{500}{\milli\molar}) solution. The $T_1$ time increases with electrolyte exposure by a factor of $1.83\pm0.35$ with respect to water. Importantly, this behavior can be altered by either the presence or absence of water or NaCl solution. After coating the same diamond with an aluminium oxide (Al\textsubscript{2}O\textsubscript{3}) thin film (thickness $\sim \SI[detect-weight]{1}{\nano\meter}$) the $T_1$ time is unaffected by NaCl solution. (b) $T_1$ relaxation curves of water and NaCl (\SI[detect-weight]{500}{\milli\molar}) solution covering the diamond surface. Diamonds were implanted with \textsuperscript{15}N at an energy of 2.5\,keV (top) and 4\,keV (bottom) resulting in different NV center ensemble depths (d\textsubscript{NV}). While on the shallower implanted diamond the $T_1$ time increases by a factor of around two with exposure to NaCl solution, on the deeper implanted diamond the effect is strongly reduced. Experiments are performed at $f_{\text{NV}}=\SI[detect-weight]{1.88}{\giga\hertz}$.}
    \label{fig:Figure3}
\end{figure}

Because of the surprising observation, that the $T_1$ time increases with diamagnetic electrolyte solutions compared to water covering the diamond, the next experiments concentrate on the mechanism behind this effect. Therefore, we probe the reversibility and passivation of the effect along with the sensor's response in dependence of its implantation depth. First, we evaluate if the extension of the $T_1$ relaxation time is a reversible process by exposing an oxidized diamond alternatingly to water and NaCl (\SI{500}{\milli\molar}) solution. Thereby, we show that the $T_1$ relaxation time is altered from ``short" in case of water exposure to ``long" when NaCl solution covers the surface (see Figure \ref{fig:Figure3}a in green). Alternating between water and NaCl solution demonstrates a $1.83\pm0.35$ fold increase of the $T_1$ time with electrolyte exposure on the oxidized diamond. Note that preparing the diamond surface with an alternative oxidizing reaction (Fenton reaction)\cite{Martin2009} leads to similar results to those with the triacid clean procedure (see Supplementary Note 5). In a next step, we examine if the effect is specific to the oxidized diamond surface. Therefore, the formerly oxidized diamond is coated with an aluminium oxide (Al\textsubscript{2}O\textsubscript{3}) thin film (thickness $\sim$ \SI{1}{\nano\meter})\cite{Liu2022} prepared by Atomic Layer Deposition (ALD). The aluminium oxide thin film ensures a controllable and uniform surface termination with hydroxyl groups. We repeat the previous experiment, but this time the $T_1$ relaxation time remains unaffected by the NaCl solution (see Figure \ref{fig:Figure3}a in black).

Additionally, we investigate if the extent of the electrolyte's effect is dependent on the depth of the embedded NV center ensemble. Therefore, we prepare two diamonds with \textsuperscript{15}N implantation energies of 2.5 and 4\,keV  with the triacid clean procedure described before and probe them with NV relaxometry. Near-surface NV centers implanted with an energy of 2.5\,keV are mainly distributed within a depth of $\sim$ \SI{5}{\nano\meter} below the surface, while ensembles created with 4\,keV \textsuperscript{15}N are located $\sim$ \SI{12}{\nano\meter} beneath the surface.\cite{JacobHenshaw} Figure \ref{fig:Figure3}b shows a significantly larger effect of the electrolyte on the relaxation time of the shallow implanted NV-diamond with respect to the deeper one, although a $T_1$ time extension is still detectable in the latter case (see also Supplementary Note 6). Importantly, while the effect with the $\sim \SI{1}{\nano\meter}$ thick aluminium oxide layer is completely passivated, a $T_1$ time increase can still be observed on the oxidized diamond when the sensor is $\sim$ 5 to 10 \si{\nano\meter} further away from the electrolyte. Note that the same measurements conducted with LiCl (\SI{500}{\milli\molar}) solution lead to similar results (see Supplementary Note 6). 

From these experiments we conclude that the extension of the $T_1$ relaxation time is a reversible and interfacial process which is dependent on the distance of the sensor to the sample.

Additionally, NV charge state alterations or changes in NV-dephasing and NV-coherence are not observed in our experiments (see Supplementary Note 7).
\FloatBarrier

\subsection{Probing Magnetic and Electric Noise Contributions}
\begin{figure} [htbp]
    \centering
    \includegraphics{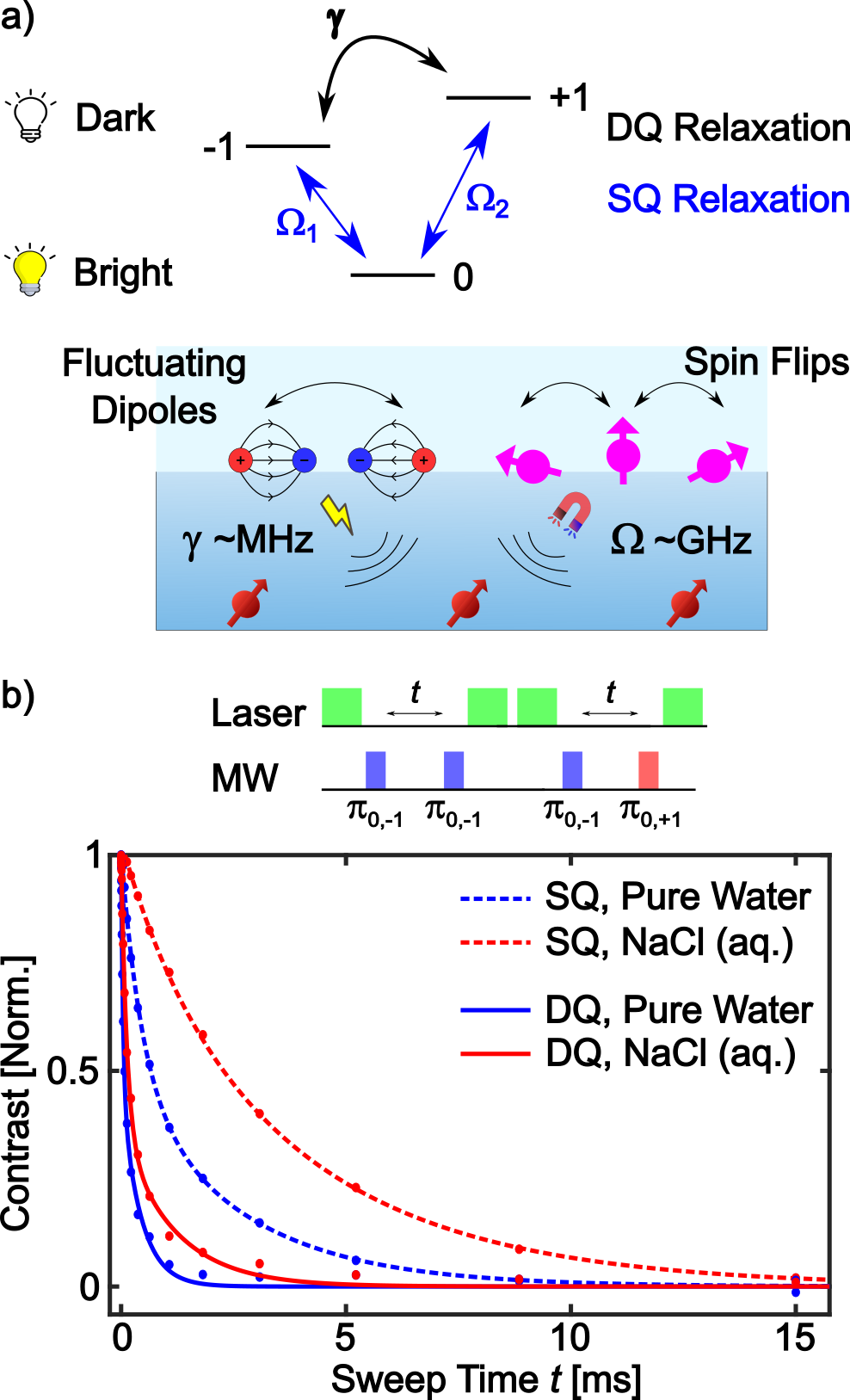}
    \caption{Single quantum (SQ) and double quantum (DQ) relaxation experiments. (a) Energy level scheme of the NV center ground state transitions. SQ transitions ($\Delta m_{s}=\pm 1$) with relaxation rates $\Omega$ are susceptible to magnetic noise. The DQ relaxation ($\Delta m_{s}=\pm 2$) with relaxation rate $\gamma$ is magnetically forbidden but susceptible to electric noise.\cite{Myers2017} (b) Top: DQ pulse sequence. Bottom: SQ and DQ relaxation curves of water and NaCl (\SI[detect-weight]{500}{\milli\molar}) solution covering the diamond. Experiments are performed at $B_0$\,=\,\SI[detect-weight]{15}{\Gauss}, where the NV\textsubscript{0,-1} transition is at $f_{\text{NV}}=\SI[detect-weight]{2.83}{\giga\hertz}$ (corresponding to a DQ transition frequency of \SI[detect-weight]{80}{\mega\hertz}). $T_{1,\text{SQ}}$ increases by a factor of $1.52 \pm 0.16$ and $T_{1,\text{DQ}}$ by a factor of $2.84 \pm 0.31$ compared to water when the diamond is exposed to NaCl solution.}
    \label{fig:Figure4}
\end{figure}

In the next set of experiments we investigate if the $T_1$ relaxation time increase originates from a reduction of electric and/or magnetic field noise. Since we are dealing with electrolytes dissolved in water, it is particularly interesting to explore the influence that charged ions and their randomly fluctuating electric fields might have on the $T_1$ relaxation time of the near-surface NV-ensembles. Whereas the typical relaxometry experiments use single quantum (SQ) transitions to probe magnetic field noise (as in the experiments from the previous sections), double quantum (DQ) transitions are influenced by electric field noise.\cite{Myers2017} For these transitions, the full NV center ground state ($S=1$) is considered, where an additional relaxation pathway between $m_{s}=-1 \leftrightarrow m_{s}=+1$ with $\Delta m_{s}=2$ (see Figure \ref{fig:Figure4}a) becomes accessible: the DQ transition. Regarding the NV center as a ``qutrit" rather than a qubit allows to probe the effect of the diamagnetic electrolyte solution on both electric and magnetic field noise at the same time.
\FloatBarrier

SQ and DQ relaxometry measurements on the system water/NaCl (\SI{500}{\milli\molar}) solution reveal that the diamagnetic electrolyte has an effect on both relaxation channels, \textit {i.e.}, it reduces magnetic as well as electric field fluctuations (see Figure \ref{fig:Figure4}b). Here, two different $T_1$ times can be defined: $T_{1,\text{SQ}}$ for the relaxation time in the SQ and $T_{1,\text{DQ}}$ in the DQ channel. An increase of $T_{1,\text{SQ}}$ by a factor of $1.52 \pm 0.16$ and an increase of $T_{1,\text{DQ}}$ by a factor of $2.84 \pm 0.31$ can be measured in presence of NaCl solution with respect to water (see also Supplementary Note 8). Thus, diamagnetic electrolytes reduce both -- electric and magnetic -- noise at the diamond surface. Importantly, when we repeat the same experiments with paramagnetic MnCl\textsubscript{2}, only $T_{1,\text{SQ}}$ reduces by 80\%, whereas the DQ transition remains unaffected compared to water (see Supplementary Note 8). This indicates an exclusive impact of the paramagnetic electrolyte on magnetic field noise and could provide a possible pathway to distinguish para- from diamagnetic ions in solution.

Additionally, we can exclude an influence of diamagnetic NaCl solution on the static electric field environment by performing zero field ESR Measurements (see Supplementary Note 8).

\subsection{Influence of Electrolytes on Surface Dark Spins: DEER Experiments}
\begin{figure} [htbp]
    \centering
    \includegraphics{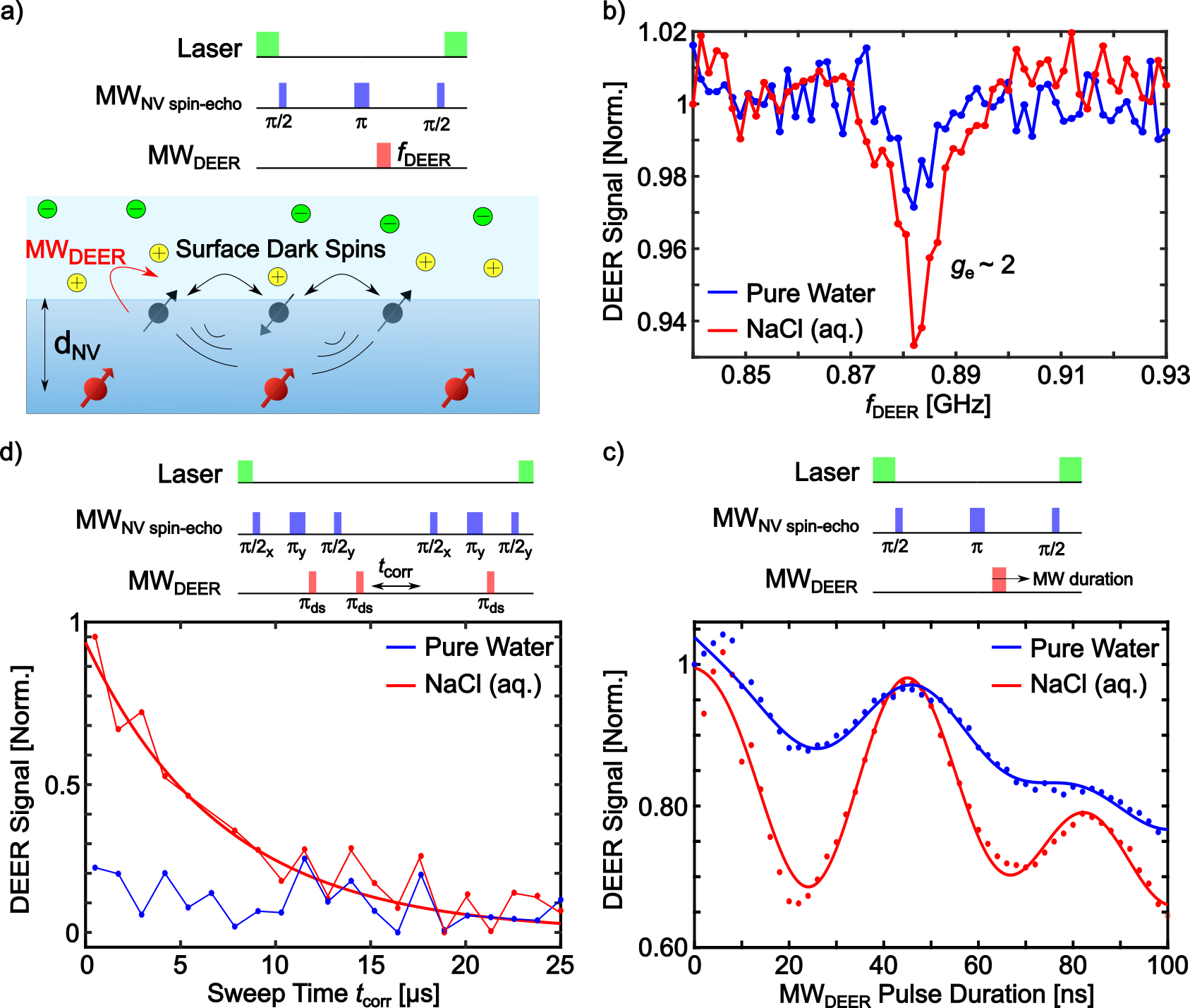}
    \caption{NV-DEER (double electron electron resonance) experiments probing the response of surface dark spins to electrolyte exposure. (a) Pulse sequence of the DEER experiment. Sweeping the microwave frequency of the MW\textsubscript{DEER} pulse (\textit{f}\textsubscript{DEER}) while applying a spin-echo experiment on the NV center (MW\textsubscript{NV spin-echo}) allows for the detection of electronic (surface dark) spins coupled to the NV centers.\cite{Mamin2012} (b) DEER experiment with water and NaCl (\SI[detect-weight]{500}{\milli\molar}) solution covering the diamond surface. A pronounced dip in the DEER spectrum appears at around \SI[detect-weight]{0.887}{\giga\hertz} (where $g_e \sim 2$) when the diamond is exposed to NaCl solution. The dip gets drastically reduced when water covers the diamond surface. (c) Top: Pulse sequence of the DEER-Rabi experiment. Bottom: When the pulse duration of the microwave drive (MW\textsubscript{DEER}) at the surface dark spins' resonance frequency is swept during the spin-echo, DEER-Rabi oscillations can be observed.\cite{Bluvstein2019} While the $\pi_{\text{ds}}$-pulse lengths remain equal when water or NaCl (\SI[detect-weight]{500}{\milli\molar}) solution cover the diamond surface ($\pi_{\text{ds}}\sim\SI[detect-weight]{24}{\nano\second}$), the latter causes a $\sim$ three times more pronounced Rabi amplitude. (d) Top: Pulse sequence of the DEER-$T_1$ experiment. Bottom: Varying the correlation time (\textit{t}\textsubscript{corr}) between two subsequent DEER segments,\cite{Sushkov2014c} shows a surface dark spin relaxation with $T_{1,\text{ds}}$\,=\,$7.20\pm1.10\,\si[detect-weight]{\micro\second}$ in case of NaCl exposure. In contrast, no significant relaxation decay can be observed when water covers the diamond surface. DEER experiments are performed at $f_{\text{NV}}=\SI[detect-weight]{1.98}{\giga\hertz}$.}
    \label{fig:Figure5}
\end{figure}

So far, we have focused on the direct influence of electrolytes on the NV centers. However, the diamond as the NV center's host material provides various surface dark spins, \textit {e.g.}, dangling bonds, whose response to the electrolytes is probed in the next set of experiments. Intrinsic $T_1$ times of these surface dark spins are often long ($\sim$ a few microseconds)\cite{Sushkov2014c} which allows us to probe them with NV-based DEER spectroscopy.\cite{Schlipf2017,Shi2015} Figure \ref{fig:Figure5}a shows the pulse sequence of a typical DEER experiment: A spin-echo is performed on the NV center's electronic spin (MW\textsubscript{NV spin-echo}), at the same time, in the second free precession time of the echo, an additional microwave pulse (MW\textsubscript{DEER}) is applied to drive the target electronic spins. Sweeping the MW frequency (\textit{f}\textsubscript{DEER}) flips the surface dark spins when their Larmor frequency is matched and causes a dip in the DEER signal (see Figure \ref{fig:Figure5}b). 
As shown in Figure \ref{fig:Figure5}b, a clear dip in the DEER spectrum appears when the diamond interface is covered with NaCl (\SI{500}{\milli\molar}) solution. The resonance at $f\textsubscript{{\text{DEER}}}=\SI{0.887}{\giga\hertz}$ corresponds to $g_e \sim 2$ spins and is typically assigned to dangling bonds at the diamond surface.\cite{Barry2020} Interestingly, the dip is drastically reduced when the experiment is repeated with water.  

Once the resonance condition for the $g_e \sim 2$ spins is found, coherent control of the surface dark spin state can be demonstrated. Figure \ref{fig:Figure5}c shows a DEER-Rabi experiment on the surface dark spins. Sweeping the microwave pulse duration (MW\textsubscript{DEER}) during the spin-echo causes oscillations of the defect's spin state.\cite{Bluvstein2019} While we determine equal $\pi_{\text{ds}}$ pulse lengths ($\pi_{\text{ds}}\sim\SI{24}{\nano\second}$) for water and the electrolyte, the former leads to a $\sim$ three-fold increased DEER-Rabi amplitude with respect to the latter. In the next step, we probe the surface dark spin relaxation time ($T_{1,\text{ds}}$). Figure \ref{fig:Figure5}d depicts a pulse sequence from Sushkov \textit{et al.},\cite{Sushkov2014c} where the $T_1$ time of the surface dark spins can be measured by correlating two subsequent DEER segments and varying the correlation time \textit{t}\textsubscript{corr}. Interestingly, we measure a relaxation time $T_{1,\text{ds}}$ of $7.20\pm1.10\,\si{\micro\second}$ for the surface dark spins exposed to the NaCl solution. In contrast, a clear relaxation decay cannot be observed in case of pure water covering the diamond. Although we cannot exclude that the surface dark spins vanish, the more likely case is that the dark spins act as reporter spins\cite{Sushkov2014c} experiencing the same effect of the electrolyte solution as the NV center: ``fast'' relaxation in the case of water and ``slow'' relaxation when diamagnetic electrolyte solutions cover the diamond, which is also expressed in the increased DEER signals (see Figure \ref{fig:Figure5}b-d). Accordingly, we enhance the sensitivity of our sensor by the proximity of the reporter spins to the electrolyte solutions.\cite{Sushkov2014c,Zhang2023}
\FloatBarrier

\section{Theoretical Modeling}
Our experimental results show an influence of diamagnetic electrolyte solutions on near-surface spin defects in diamond where electric as well as magnetic noise is suppressed resulting in an increase of the $T_1$ relaxation time. To further study this surprising effect computational modeling is used. Here, as a working assumption, we focus on charge fluctuations within the diamond lattice. We note, that further processes such as proton hopping at the interface or water and ion dynamics within the electric double layer can also play a role but have not been treated herein. 

\begin{figure}
    \centering
    \includegraphics{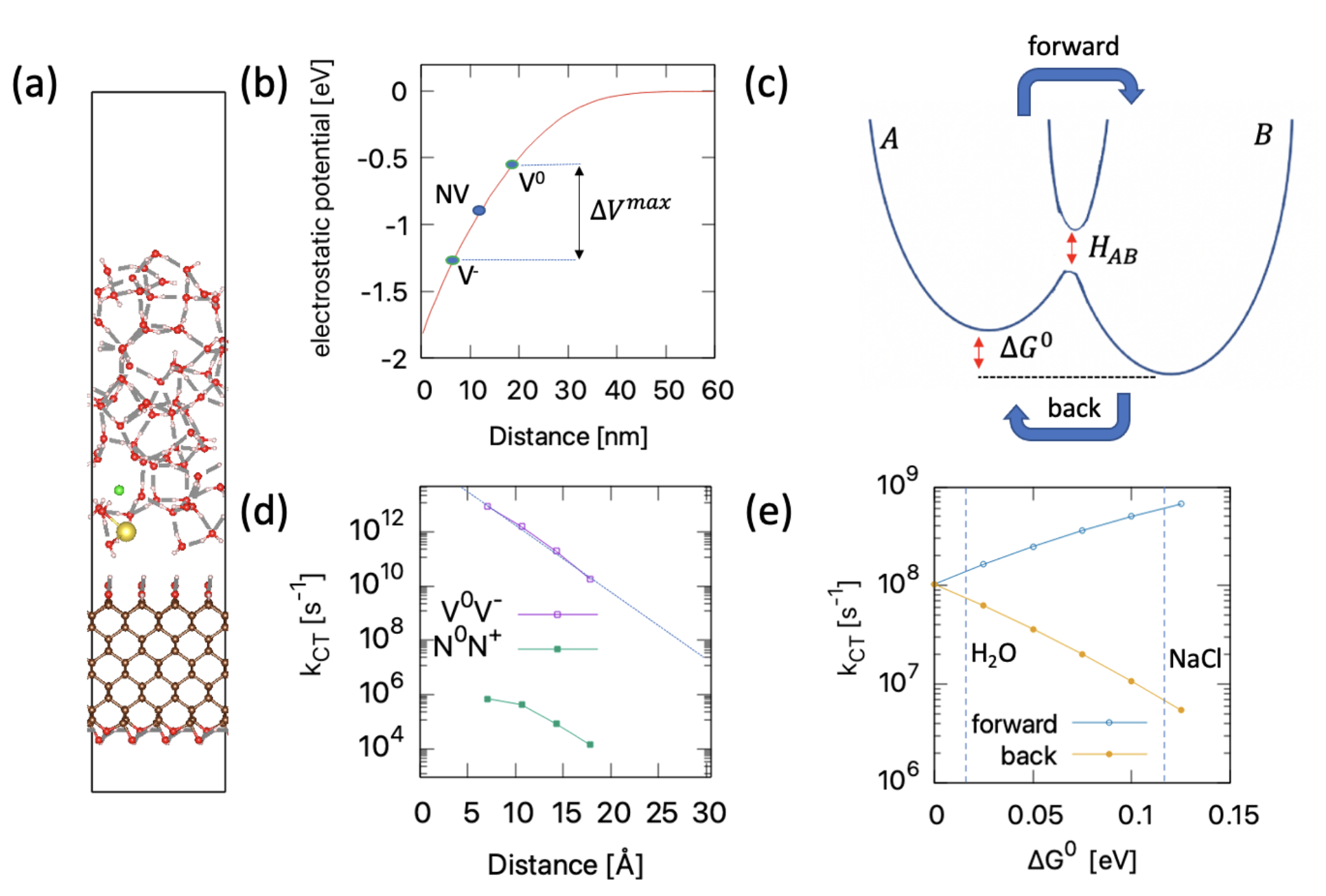}
    \caption{\textit{Ab initio} MD simulations of the diamond/electrolyte interface. (a) Representative snapshot of the diamond/electrolyte interface from the \textit{ab initio} MD simulations. Color code: C (grey), O (red), H (white), Na\textsuperscript{+} (yellow), Cl\textsuperscript{-} (green). (b) Variations of the electrostatic potential in a semi-infinite diamond due to the arrangement of NaCl at the interface. Inset: Electrostatic potential ($\Delta V_{\text{max}}$) for a pair of defects in \SI{3}{\nano\meter} distance to each other surrounding a $\sim$ \SI[detect-weight]{5}{\nano\meter} deep NV center. (c) Schematic representation of a continuous charge hopping between two defects around the NV center. $H_{AB}$ is the transfer integral, $\Delta G^0$ is calculated as $\Delta V_{\text{max}}/2$ from (b). (d) Calculated rate constants for pairs of substitutional nitrogen and vacancy defects as a function of distance between the defects. (e) Rate constants for the forward and backward electron transfer between a pair of vacancy defects as a function of bias due to the interfacial band bending. Vertical lines refer to an estimated difference in the effects by replacing the electrolyte with pure water, considering a pair of defects and an NV center in a configuration from the inset in (b).}
    \label{fig:Figure6}
\end{figure}

To this end, we model an interface between a slab of diamond and a thin layer of water subsequently enriched by Na\textsuperscript{+} and Cl\textsuperscript{-} ions (see Methods for detail). Then, we probe the interfacial structure and vacuum level shifts (VLS) based on the configurations obtained from the \textit{ab initio} molecular dynamics (MD) (see Figure \ref{fig:Figure6}a). The calculated alignment of the electronic levels of water and the model diamond surface is shown in Figure \ref{fig:FigureS12}. Here, a mismatch of the chemical potentials (defined as a center of the band gap) promotes an electron leakage from the diamond surface towards the water. The resulting redistribution of charges leads to the development of an electric field, that further rearranges the charged solvated Na\textsuperscript{+} and Cl\textsuperscript{-} ions. The large positive VLS of 1.1\,eV (see Figure \ref{fig:FigureS12}a) causes the ions to rearrange with the direction of the field, facilitating the effect of band bending. By adding a carboxyl group, we observe a stabilization of the downwards band bending relative to the case of the model diamond surface in water. However, in both cases, we obtain a broad distribution of surface dipoles, owing to the complexity of ion dynamics within the Stern layer. By contrast, a sharp distribution of the dipole moments is observed between a dissociated carboxyl group and a solvated Na\textsuperscript{+} ion nearby. This stable configuration gives rise to a large VLS of $\sim$ -1.9\,eV. This value is further used to trace the evolution of the electrostatic potential at the microscopic level. More specifically, we set it as a boundary condition for solving the Poisson equation to access the modifications of the potential inside a semi-infinite diamond slab. As shown in Figure \ref{fig:Figure6}b, we observe that the interfacial region of $\sim$ \SI{40}{\nano\meter} is affected by the respective readjustments of the charges, resulting in a rapid decay of the potential near the interface and a slow saturation towards the bulk.

To establish a relation between the band bending and the noise reduction, we consider the electric and magnetic fluctuations caused by a pair of active defects around the NV centers. The charge transfer process leads to a continuous change in the charge/spin state of the nearby defects, which can affect the relaxation and coherence time of the NV center when the rate approaches the timescale of the quantum sensing experiment. In the Marcus theory, such fluctuations are described as a sequence of thermally activated hopping events, whilst the rate constants are determined from the distance-dependent coupling parameters and the required structural reorganizations (see Figure \ref{fig:Figure6}c). The dipoles at the diamond/solvent interface affect this equilibrium by altering the onsite Gibbs energy term with a contribution from the electrostatic potential ($\Delta V$). As shown in Figures \ref{fig:Figure6}c and \ref{fig:Figure6}e, the band bending accelerates a forward charge transfer process (until reaching the Marcus inverted region), but at the same time, the magnitude of the back charge transfer (BCT) rate drops exponentially. Hence, regardless of the defect type, large interfacial band bending can lead to a dynamical trapping of the charges around a site with the lower Gibbs energy. For a numerical validation, we focus on the electron fluctuations between a pair of carbon vacancies as well as on the hole fluctuations between two substitutional nitrogen defects. We note that carbon vacancies are often generated by the implantation and irradiation techniques used to create the NV centers, and substitutional N defects are present around NV centers to stabilize the negative charge state of the NV center. After determining the relevant parameters in the periodic supercells (see Methods for detail), we first analyze the possible contribution of each charge transfer reaction to the electric noise. To this end, we compare the fluctuation rates, calculated for both defect pairs in a bulk-like environment ($\Delta V=0$). As shown in Figure \ref{fig:Figure6}d, the charge fluctuation rates for a pair of nitrogen defects is remarkably smaller than for a vacancy pair. This difference is attributed to a formation of an energetically unfavorable N\textsuperscript{0} configuration, that hinders the charge fluctuations by large structural reorganizations ($\lambda_\text{reorg}=1.89\,\text{eV}$). Hence, substitutional N\textsuperscript{0} pairs are unlikely to be the origin of electric and magnetic noise, affecting the $T_1$ time due to charge fluctuation. By contrast, owing to a rather modest $\lambda_\text{reorg}$ of 0.28\,eV, a vacancy pair gives rise to noise in a broad frequency range, where the respective rate constants are controlled by the separation between the active sites.

The relevant distance between a pair of vacancies is readily obtained from an exponential fit of the rate constants in Figure \ref{fig:Figure6}d. More specifically, we find the charge fluctuation rates, which would be relevant for an influence on the $T_1$ time ($\sim$ 0.1\,\si{\giga\hertz}) at a separation of $\sim$ \SI{3}{\nano\meter}. The molecular dynamics simulations performed by Fávaro de Oliveira \textit{et al.} show that such a high local density of vacancies around the region of an NV center is achieved during the nitrogen implantation due to the cascade process of the ``kick-out'' mechanism.\cite{FavarodeOliveira2017} Using the variations of electrostatic potential from Figure \ref{fig:Figure6}b for the implantation depths of 5 and 12 nm, we calculate a contribution to the onsite Gibbs energies by 0.125 and 0.075\,eV, respectively (see Methods for details). As shown in Figure \ref{fig:Figure6}e, the BCT rate at 12\,\si{\nano\meter} reduces by a factor of $\sim$\ 7 relative to the case of $\Delta V=0$. Moreover, in agreement with our experimental results, the faster changes in the potential at \SI{5}{\nano\meter} enhances the reduction factor to $\sim$ 19 for the shallower NV centers. Relative to pure water, the BCT rates decrease by factors of $\sim$ 10 and 3.5 for the depths of 5\,\si{\nano\meter} and 12\,\si{\nano\meter}, respectively. Furthermore, the proposed mechanism is consistent with the experimental results on dark spins (see Figure \ref{fig:Figure5}). Interestingly, our calculations point to an even larger decrease of the BCT rate at smaller distances from the interface which should translate to an even larger sensitivity. Given the favorable downwards band bending, these results call for a further optimization of the implantation parameters as well as the surface structure to fully exploit the extension of the $T_1$ time by diamagnetic electrolyte solutions. We note that a similar impact of paramagnetic ions on the band bending is very likely but obscured by the strong magnetic interaction of the NV center with unpaired electrons leading to an overall decrease in the $T_1$ time. 

\FloatBarrier
\section{Conclusion and Outlook}
We report on the effect of diamagnetic electrolyte solutions on highly dense near-surface spin defects in oxidized diamonds. Surprisingly, we observe that diamagnetic ions increase the $T_1$ relaxation time of NV centers. We demonstrate that this effect is reversible, surface sensitive and responsive to millimolar concentrations. We also find that interfacial spin defects are sensitive to diamagnetic species, anticipating their possible use as reporter spins for future optimization. Furthermore, we investigate the underlying mechanism by single and double quantum NV relaxometry experiments in combination with \textit{ab initio} simulations. We propose that ions at the interface stabilize charge fluctuations between pairs of carbon vacancies and alike deep defects surrounding the NV centers. This reduces magnetic as well as electric noise at the diamond interface by a dynamical trapping of mobile electrons to a site with lower Gibbs energy. 

Our findings mark an important step towards understanding the complex processes at the solid/liquid interface of an oxidized NV-diamond. Further studies will be necessary to obtain a complete understanding of the $T_1$ time increase by electrolyte solutions. For instance, electric noise at the surface could be probed with more advanced sensing schemes\cite{Li2020,Cheng2023}.
Further exploration of diamond surface terminations, including nitrogen-termination\cite{Kawai2019} and highly ordered oxygen-termination\cite{Sangtawesin2019}, could provide insight into how chemical groups interact with the electrolyte. However, altering the surface termination may also modify the properties of the NV-diamond (such as band bending,\cite{Kaviani2014a} electron affinity,\cite{Cui2013} NV charge state,\cite{Kawai2019} surface dark spins,\cite{Janitz2022} \textit{etc.}), which can, in turn, impact the NV center's relaxation. Thus, careful consideration must be taken regarding the $T_1$ time behavior. In addition, different implantation parameters or NV creation techniques, such as helium ion implantation\cite{Huang2013} and nitrogen delta-doping,\cite{Ohno2012} could shed light on the role of implantation-induced (vacancy) defects on the $T_1$ time. Furthermore, we encourage similar experiments with high spatial resolution or single NV centers\cite{Zheng2022,Dwyer2022} to correlate the effect of electrolytes with the NV center's local environment. Similarly, probing the influence of electrolytes on NV centers in nanodiamonds\cite{Schirhagl2014a} should be a crucial objective.   

In addition to advancing our fundamental understanding of interfacial processes on spin defects in diamond, we anticipate a range of potential applications, such as electrolyte sensing in cell biology,\cite{Sigaeva2022} neuroscience,\cite{Hall2010b} or electrochemistry.\cite{Castelletto2023} 

\section{Methods}

\subsection{Sample Preparation}
Two $2\times2\times0.5\,\si{\milli\meter}$ electronic grade diamond samples (natural \textsuperscript{13}C abundance,  Element Six) were implanted with \textsuperscript{15}N at an energy of 2.5\,keV and 4\,keV, an off-axis tilt of \SI{7}{\degree} and a fluence of $2\times10^{12}\,\si{\per\square\centi\meter}$ by Innovion and annealed according to Bucher \textit{et al}.\cite{Bucher2019} Before experiments are conducted, the diamonds are cleaned with a triacid cleaning protocol according to Brown \textit{et al.}:\cite{Brown2019} Samples are boiled in equal parts of sulfuric, nitric and perchloric acid at a temperature of \SI{280}{\celsius} for two hours. This cleaning procedure is also applied before the deposition of aluminium oxide (\ce{Al2O3}) on the diamond. 

\subsection{Preparation of Electrolyte Solutions}
For the measurements where pure water is used, deionized water with a resistivity of \SI{18.2}{\mega\ohm}$\cdot$\si{\centi\meter} at \SI{25}{\celsius} (Merck Millipore) is utilized.

Sodium chloride (NaCl, Merck 106404) is prepared in a \SI{1}{\molar} stock solution, where NaCl is dissolved in deionized water. Before the experiments, NaCl is diluted from the stock solution in steps of 1:10. The other salt solutions used within this work are prepared in the same manner.

\subsection{Atomic Layer Deposition (ALD)}
The 2.5\,keV \textsuperscript{15}N implanted diamond is coated with an aluminium oxide (\ce{Al2O3}) thin film by ALD according to Liu \textit{et al.}\cite{Liu2022} The deposition includes 10 cycles of alternated sample exposure to trimethyl aluminium (TMA) and \ce{H2O}. This procedure results in a film thickness of $\sim$ \SI{1}{\nano\meter} and ensures surface termination with hydroxyl groups by exposing the diamond to a remote oxygen plasma within the ALD system.\cite{Liu2022,Henning2021} The \ce{Al2O3} layer can be removed from the diamond surface by soaking the sample overnight in 5\% NaOH solution. 

\subsection{Experimental Setup}
The quantum sensing setup is based on a modified version of the experiment described in Bucher \textit{et al.}\cite{Bucher2019} Before experiments are performed, the diamond is glued to a thin glass cover slide (48393026, VWR) together with a microfluidic device that encloses the diamond edges and covers its surface, such that a volume of $\sim$ \SI{0.60}{\micro\liter} of the sample liquid can be applied in a controllable way.\cite{Allert2022b} On the other side of the cover slide a \SI{6}{\milli\meter} diameter glass hemisphere (TECHSPEC\textsuperscript{\textregistered} N-BK7 Half-Ball Lenses, Edmund Optics) is glued, in order to improve the fluorescence light collection efficiency. The glass cover slide is then fixed on a \SI{30}{\milli\meter} cage plate (CP4S, Thorlabs). This whole assembly is then positioned between two permanent magnets, that are rotated and tilted in order to align the $B_0$ field with one of the four possible NV center orientations. The distance between the two magnets can be adjusted in order to correspond to the working magnetic field strengths $B_0$ (in this work: 15, 316, 352 and 978 \si{\Gauss}). Initialization of the NV-ensemble is realized with a \SI{532}{\nano\meter} laser (Verdi G5, Coherent) with a power of $\sim$ \SI{250}{\milli\watt} (CW) after the acousto-optic modulator (AOM). The laser light is focused on the diamond by a Plano-Convex Lens (LA 1986-A-M, Thorlabs) in a total internal reflection geometry. Laser pulses are regulated by an AOM (Gooch and Housego, model 3250-220) with pulse durations of \SI{5}{\micro\second}. Photoluminescence (PL) is collected and focused on a large area photodiode (OE-300-SI-10, Femto Messtechnik GmbH, Berlin, Germany) by two condenser lenses (ACL25416U-B, Thorlabs). The excitation light is filtered by a long-pass optical filter (Edge Basic 647 Long Wave Pass, Semrock) placed between the bottom condenser lens and the photodiode. The output voltage of the photodiode is digitized with a data acquisition unit (USB-6229 DAQ, National Instruments). A \SI{500}{\mega\hertz} PulseBlaster card (ESR-Pro-II,
Spincore) is utilized to trigger and to time the microwave and light pulses used for quantum control of the NV centers. The microwave frequencies are produced by a signal source (SynthHD, Windfreak Technologies, LLC.). Microwave phase control is obtained by a combination of a phase-shifting splitter (ZX10Q-2-27-S+, Mini-Circuits), two switches (ZASWA-2-50dRA+, Mini-Circuits) and a combiner (ZX10-2-42-S+, Mini-Circuits). The amplified microwave pulses (ZHL-16W-43-S+, Mini-Circuits) are delivered by a homebuilt microwave loop on top of the microfluidic chip.\cite{Bucher2019} The electron spin resonance (ESR) frequency is used to determine the magnetic field strength $B_0$ as well as the NV\textsubscript{0,-1} resonance frequency \textit{f}\textsubscript{NV}. 

\subsection{\boldmath{$T_1$} Relaxometry Experiments (Single and Double Quantum)}
Single quantum (SQ) relaxometry experiments: To obtain a signal-to-noise ratio (SNR) as shown in Figure \ref{fig:Figure1}b the sequence is repeated 5,000 times for every data point. Each experiment consists of 31 data points measured in a logarithmic increasing sweep time \textit{t} to guarantee more sampling points at short times \textit{t}. Note that these parameters are also used for double quantum (DQ) relaxometry. For normalization and noise cancellation, the second half of the sequence contains a MW $\pi_\text{0,-1}$-pulse, where the subscripts 0 and -1 indicate the transition of the spin state between $m_s=0$ and $m_s=-1$.\cite{Bucher2019} The spectra are then plotted as the measurement result of the first half divided by the result of the second half of the sequence.

Double quantum (DQ) relaxometry experiments: For a detailed discussion of the DQ relaxation and pulse sequence, the reader is referred to Myers \textit{et al.}\cite{Myers2017} In short, the DQ pulse sequence (see inset of Figure \ref{fig:Figure4}b) consists of two consecutive measurements where MW $\pi$-pulses are used to control spin state initialization and readout. In both halves of the sequence the NV center is initialized in $m_{s}=-1$. After a sweep time \textit{t} the spin state population of either $m_{s}=-1$ (in the first part) or $m_{s}=+1$ (in the second part) is read out. Dividing the second by the first part yields a population ratio of the two states. 

\subsection{Sensitivity of $T_1$ Relaxometry on Electrolytes}
Experiments to determine the sensitivity of $T_1$ relaxometry measurements on para- and diamagnetic electrolytes are conducted for MnCl\textsubscript{2} and NaCl solutions using the SQ relaxometry pulse sequence. Probing each concentration results in a relaxation curve of which the $T_1$ time is determined. The $T_1$ time is then normalized to the one of water covering the diamond. Before probing any electrolyte concentration, we wash the microfluidic device with water to ensure equal starting conditions, \textit {i.e.}, a constant $T_1$ time for water covering the diamond. We perform each series three times resulting in a mean $T_1$ value for each concentration (see also Supplementary Note 4). Figure \ref{fig:Figure2} in the main text shows the mean (normalized) $T_1$ time along with the standard deviation. 
\subsection{DEER Measurements}
DEER spectra (see Figure \ref{fig:Figure5}b) are recorded by performing a spin-echo sequence on the NV center spins with a free evolution time of \SI{1}{\micro\second}. The duration of the MW-pulse (MW\textsubscript{DEER}) applied to the surface dark spins is set to \SI{200}{\nano\second} and the driving frequency (\textit{f}\textsubscript{DEER}) is swept over \SI{90}{\mega\hertz} (from $\textit{f}\textsubscript{DEER}=0.84$ to \SI{0.93}{\giga\hertz}). To obtain an SNR as shown in Figure \ref{fig:Figure5}b the sequence is repeated 10,000 times for every data point. Each experiment consists of 67 data points in equally separated frequency steps and this whole experiment is repeated four times. Referencing for noise cancellation is achieved by alternating the last MW-pulse of the spin-echo sequence from $\pi/2$ to $3/2\pi$. 

Once the resonance condition for $g\textsubscript{e}=2$ is found, DEER-Rabi experiments on the surface dark spins are performed by sweeping the MW-pulse duration (MW\textsubscript{DEER}) during the NV spin-echo (see Figure \ref{fig:Figure5}c) as described above. The sequence is repeated 10,000 times for every data point. Each experiment consists of 101 equally spaced data points and this whole experiment is repeated ten times. To account for MW (MW\textsubscript{DEER}) noise, the same procedure is repeated \SI{20}{\mega\hertz} off the resonance condition. The outcome of both on- and off-resonant measurements are subtracted resulting in the spectra shown in Figure \ref{fig:Figure5}c. After that, measurements of the surface dark spin population relaxation are carried out according to Sushkov \textit{et al.} with a $\pi_{\text{ds}}$-pulse length of \SI{24}{\nano\second}.\cite{Sushkov2014c} The sequence shown in Figure \ref{fig:Figure5}d is repeated 10,000 times for every data point. Each experiment consists of 21 data points in equally separated time steps. This whole experiment is then repeated 50 times. Background subtraction is achieved by performing the experiment in the same procedure without the additional MW drive (MW\textsubscript{DEER}). Subtracting the outcome of both MW-on and MW-off measurements then yields the spectra shown in Figure \ref{fig:Figure5}d.

\subsection{Simulation of the Diamond/Water Interface}
In our simulations, we use a slab of a model diamond surface with hydrogen, hydroxyl, and ether surface terminations (see Figure \ref{fig:FigureS12}d). It is a symmetric (100) surface of $\sim$\,1.4\,\si{\nano\meter} with a $2\times1$ surface reconstruction pattern, exhibiting a positive electron affinity and no surface states inside the band gap\cite{Kaviani2014a}. The water layer on top of the diamond (thickness $\sim \SI{2}{\nano\meter}$) was constructed as follows. First, we equilibrate 74 water molecules with the classical molecular dynamics (MD) for \SI{5}{\nano\second} in a simulation box of commensurate lateral size with the diamond slab. These calculations are done with the GROMACS software in the canonical NVT ensemble,\cite{Berendsen1995} using the GROMOS 54A7 force field.\cite{Schmid2011} After that, we superimpose the water box and the diamond surface and allow for an additional equilibration step of 10\,ps with the \textit{ab initio} MD, as implemented in the VASP package.\cite{Kresse1996} We also incorporate $\sim$\,1.9\,nm of vacuum together with a dipole correction scheme to eliminate the interaction with the periodic images. This yields the simulation supercell of $1.0097\times 1.0097\times5.3\,\si{\cubic\nano\meter}$, which is further used in the \textit{ab initio} MD calculations. \textit{Ab initio} calculations are performed using the PBE functional\cite{Perdew1996} in conjunction with the D2 dispersion correction, using a projector augmented wave method with the kinetic energy cutoff of 370\,eV. We note that the PBE functional provides semi-quantitative results for the electronic structure but is able to accurately yield the trends in the change of the electronic structure upon different surface terminations and environments of diamond. Further, we note that we focus on the difference in the electrostatic environment due to the interaction of the electrolyte with the surface groups, assuming no change in the microstructure of the carbon layer.

\subsection{Charge Transfer Rates}
We calculate the charge transfer rate with an expression from the Marcus theory,\cite{Marcus1956} given as:
\begin{equation*}
    k_{CT}=\frac{2\pi}{\hbar}|H_{AB}|^2\frac{1}{\sqrt{4\pi\lambda k_{B}T}}\exp{\Bigl(-\frac{(\lambda+\Delta G)^2}{4\pi\lambda k_{B}T}}\Bigr)
\end{equation*}
where $H_{AB}$ is the transfer integral, $\lambda$ the reorganization energy, $\Delta G$ the Gibbs energy difference due to an external field, $k_{B}$ the Boltzmann constant, $\hbar$ the reduced Planck constant, and $T$ the temperature. The reorganization energy is determined for a single defect (either a carbon vacancy or a substitutional nitrogen) in a 1000-carbon supercell. For computing the transfer integrals as a function of distance, we use diamond supercells of different sizes, varying between 64 and 1000-carbon atoms. The reorganization energies are calculated by the four-point scheme, while the transfer integrals are estimated at a high symmetry configuration as 1/4 of the bandwidth along the $\Gamma$-X direction. The contribution to the Gibbs energy is computed by solving the one dimensional Poisson equation given the experimental depth of the NV center.\cite{Broadway2018} Noteworthy, the effect from the band bending is governed by the orientation of defect pairs relative to the direction of the electric field. At a reference depth of the NV center, the maximum strength, corresponding to a change in the electrostatic potential ($\Delta V_{\text{max}}$), is reached in a parallel configuration, whilst the effect is quenched towards the orthogonal arrangement. Considering a uniform distribution of the defects in our samples, we compute an expectation value of $\Delta V$ as $\Delta V_{\text{max}}/2$. 

\begin{acknowledgement}
This study was funded by the Deutsche Forschungsgemeinschaft (DFG, German Research Foundation) - 412351169 within the Emmy Noether program. R. Rizzato acknowledges support from the DFG Walter Benjamin Programme (Project RI 3319/1-1). D. Bucher acknowledges support from the DFG under Germany’s Excellence Strategy—EXC 2089/1—390776260 and the EXC-2111 390814868. M. Brandt acknowledges support by BMBF through project epiNV (13N15702) and the Munich Center for Quantum Science and Technology (MCQST, EXC-2111) and by Bayerisches Staatsministerium f\"ur Wissenschaft und Kunst through project IQSense via the Munich Quantum Valley (MQV). A. Gali acknowledges the Hungarian NKFIH grant No. KKP129866 of the National Excellence Program of Quantum-coherent materials project, the support for the Quantum Information National Laboratory from the Ministry of Culture and Innovation of Hungary (NKFIH grant No. 2022-2.1.1-NL-2022-00004), the EU EIC Pathfinder project "QuMicro" (grant No. 101046911) and the EU QuantERA for the project MAESTRO. We acknowledge KIF\"U for awarding us access to computational resources based in Hungary.
\end{acknowledgement}
\section{Abbreviations}
NV, nitrogen-vacancy; NMR, nuclear magnetic resonance; ESR, electron spin resonance; ODMR, optically detected magnetic resonance; DEER, double electron electron resonance; MW, microwave; ALD, atomic layer deposition; SQ, single quantum; DQ, double quantum; VLS, vacuum level shift; MD, molecular dynamics; BCT, back charge transfer; TMA, trimethyl aluminium; CW, continuous wave; AOM, acousto-optic modulator; PL, photoluminescence; SNR, signal-to-noise ratio; PBE, Perdew-Burke-Ernzerhof; DFT, density functional theory.
\section{Author Contributions}
D. Bucher, R. Rizzato and F. Freire-Moschovitis discovered the effect of diamagnetic electrolytes on the relaxation of NV centers. D. Bucher, R. Rizzato and F. Freire-Moschovitis designed the experiments. D. Bucher supervised the study. F. Freire-Moschovitis performed the experiments and was supported by M. Schepp for NV relaxometry. F. Freire-Moschovitis, R. Rizzato and D. Bucher analyzed the data. R. Allert built the quantum sensing setup and designed the microfluidic device. A. Gali and A. Pershin incorporated theoretical modeling and simulations. M. Brandt and L. Todenhagen helped with the charge state experiments. All authors discussed the results and contributed to the writing of the manuscript.

\section{Additional Information}
\textbf{Competing interests:} The authors declare no competing interests.
\section{Data Availability}
The data supporting our findings are available within the paper and the Supplementary Information. Additional relevant data are available from the corresponding author upon reasonable request.
\section{Code Availability}
The codes used for data acquisition and processing are available from the corresponding author upon reasonable request.

\begin{suppinfo}
Fitting details, data of additional (single and double quantum) relaxometry experiments of various electrolyte solutions and organic solvents, data of NV relaxometry measurements on differently oxidized surface terminations, data of NV charge state, coherence and dephasing measurements, data of ESR experiments at zero magnetic field, and results of DFT-PBE \textit{ab initio} molecular dynamics simulations.
\end{suppinfo}


\newpage
\section{Supporting Information: The Role of Electrolytes in the Relaxation of Near-Surface Spin Defects in Diamond}

\renewcommand{\thefigure}{S\arabic{figure}}
\setcounter{figure}{0}
\renewcommand{\thetable}{S\arabic{table}}
\setcounter{table}{0}
\subsection{Supplementary Note 1: Fitting of \boldmath{$T_1$} Single Quantum (SQ), Double Quantum (DQ) and Surface Dark Spin Relaxation Curves}
\label{SI_Note_1}
Recorded single quantum (SQ) and double quantum (DQ) relaxation curves are fitted with a biexponential function as the $T_1$ decay exhibited two components according to prior work:\cite{Steinert2013a,PeronaMartinez2020a,Li2022,Rioux2016a,Ziem2013a} 
\begin{equation*}
    C(t)=A\cdot\exp(-\frac{1}{T_{1a}}\cdot t)+(1-A)\cdot\exp(-\frac{1}{T_{1b}}\cdot t) 
\end{equation*}
where $C$ is the contrast, $A$ is the amplitude and $T_{1a} >> T_{1b}$. For completeness, relaxation times in the tables are given by both time constants. In agreement with prior work,\cite{Steinert2013a,PeronaMartinez2020a,Ziem2013a} values of $T_1$ in the main text are only considering the longer component $T_{1a}$. However, both time constants are longer in all cases where diamagnetic electrolytes are measured with NV relaxometry and compared to water (see Table \ref{tab:TableS1.2}). Errors and errorbars from SQ and DQ relaxation curves shown in tables or figures are standard deviations from the biexponential fit function or in case of the sensitivity experiments (see Figure 2 in the main text) the standard deviation from three consecutive measurements. $T_1$ time constants in the tables are given to three significant digits. In case of the $T_{1,\text{ds}}$ relaxation measurements (see Figure 5d in the main text), the relaxation curve is fitted to a single exponential decay: $C(t)=A\cdot\exp(-\frac{1}{\text{$T_{1,\text{ds}}$}}\cdot t)$.\cite{Sushkov2014c}
\subsection{Supplementary Note 2: \textit{T\textsubscript{1}} Time Constants of Measured Electrolytes and \textit{T\textsubscript{1}} Time Magnetic Field Dependence for Pure Water/NaCl (500 mM)}
\begin{figure}[htbp]
    \centering
    \includegraphics[scale=0.30]{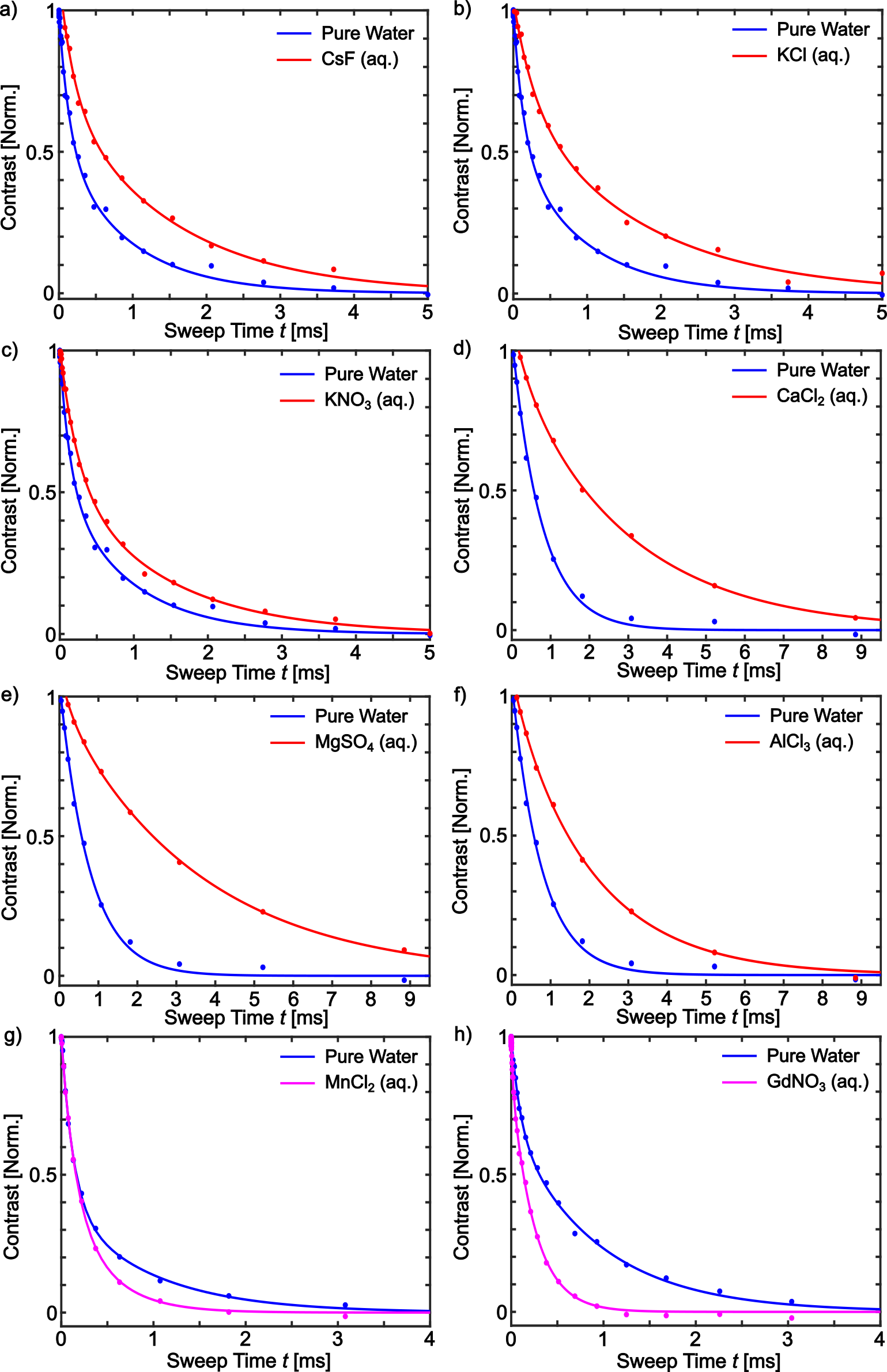}
    \caption{$T_1$ relaxation curves of water and a-f) diamagnetic electrolyte (500 mM) solutions as well as g) and h) paramagnetic electrolyte (1 \si[detect-weight]{\micro\molar}) solutions covering the diamond surface. Experiments are performed at \textit{f}\textsubscript{\text{NV}} = 1.88 GHz.}
    \label{fig:FigureS1}
\end{figure}

Table \ref{tab:TableS1.2} and Figure \ref{fig:FigureS1} show the $T_1$ time constants ($T_{1a}$ and $T_{1b}$) and $T_1$ relaxation curves of the measured electrolyte solutions in this work. Experiments are conducted with the relaxometry pulse sequence according to the main text.

 \begin{table}[htbp]
    \caption{$T_1$ time constants ($T_{1a}$ and $T_{1b}$) of water and measured diamagnetic electrolyte solutions (500 mM) as well as paramagnetic electrolyte solutions (1 \si[detect-weight]{\micro\molar}) covering the diamond surface. Experiments are performed at \textit{f}\textsubscript{\text{NV}} = 1.88 GHz.}
    \begin{tabular}{lll}
     \hline
    Electrolyte [c\,=\,\SI{500}{\milli\molar}]&$T_{1a}$ [\si{\micro\second}]&$T_{1b}$ [\si{\micro\second}]\\
    \hline
Water&$920\pm170$&$140\pm60.0$\\
CsF&$1510\pm250$&$200\pm60.0$\\
KCl&$1720\pm300$&$270\pm90.0$\\
KNO\textsubscript{3}&$1360\pm220$&$240\pm50.0$\\
LiCl&$2940\pm720$&$930\pm40.0$\\
NaCl&$1920\pm200$&$170\pm20.0$\\
CaCl\textsubscript{2}&$2920\pm260$&$460\pm190$\\
MgSO\textsubscript{4}&$3600\pm160$&$310\pm100$\\
AlCl\textsubscript{3}&$2070\pm370$&$360\pm190$\\
    \hline
    Electrolyte [c\,=\,\SI{1}{\micro\molar}]&$T_{1a}$ [\si{\micro\second}]&$T_{1b}$ [\si{\micro\second}]\\
    \hline
    MnCl\textsubscript{2}&$430\pm160$&$140\pm50.0$\\
    Gd(NO\textsubscript{3})\textsubscript{3}&$250\pm15.0$&$21\pm6.00$\\
    \hline
    \end{tabular}
    \label{tab:TableS1.2}
\end{table}

\FloatBarrier

Further measurements of water/NaCl (\SI{500}{\milli\molar}) solution are performed in different magnetic fields $B_0$ (978, 352, 15 and 0 \si{\Gauss}), \textit {i.e.}, different resonance frequencies of the NV center's $m_{s}=0 \rightarrow m_{s}=-1$ transition ($f\textsubscript{\text{NV}}=$ 0.131, 1.88, 2.83 and 2.87 \si{\giga\hertz}). Figure \ref{fig:FigureS2} shows the $T_{1a}$ time constants depending on $f\textsubscript{\text{NV}}$ (see Supplementary Note 1 for details). In Table \ref{tab:TableS1} both time constants ($T_{1a}$ and $T_{1b}$) are listed. 
\begin{figure} [htbp]
    \centering
    \includegraphics[scale=0.45]{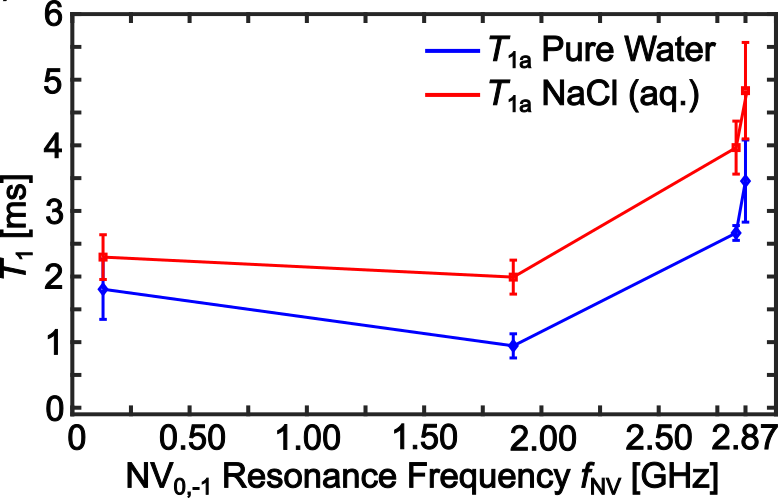}
    \caption{\textbf{$T_{1a}$ time constants for water and NaCl (500 mM) solution and their dependence on the NV\textsubscript{0,-1} resonance frequency $f\textsubscript{\text{NV}}$}.}
    \label{fig:FigureS2}
\end{figure}

\begin{table}[htbp]
   
    \caption{\textbf{$T_1$ time constants ($T_{1a}$ and $T_{1b}$) for water and NaCl (500 mM) solution covering the diamond surface depending on the NV\textsubscript{0,-1} resonance frequency $f\textsubscript{\text{NV}}$}.}
    \begin{tabular}{lll}
    \hline
    &$T_{1a}$ [\si{\micro\second}]&$T_{1b}$ [\si{\micro\second}]\\
    \hline
   $f\textsubscript{\text{NV}}$ = \SI{0.131}{\giga\hertz} \\
    Water&$1810\pm460$&$400\pm60.0$\\
    NaCl \SI{500}{\milli\molar}&$2300\pm340$&$670\pm130$\\
   \hline
   $f\textsubscript{\text{NV}}$ = \SI{1.88}{\giga\hertz} \\
    Water&$940\pm180$&$130\pm20.0$\\
    NaCl \SI{500}{\milli\molar}&$1990\pm200$&$170\pm20.0$\\
     \hline
    $f\textsubscript{\text{NV}}$ = \SI{2.83}{\giga\hertz} \\
    Water&$2660\pm110$&$510\pm80.0$\\
    NaCl \SI{500}{\milli\molar}&$3970\pm400$&$1210\pm410$\\
     \hline
    $f\textsubscript{\text{NV}}$ = \SI{2.87}{\giga\hertz} \\
    Water&$3460\pm630$&$930\pm190$\\
    NaCl \SI{500}{\milli\molar}&$4830\pm740$&$1830\pm510$\\
     \hline
    \end{tabular}
    \label{tab:TableS1}
\end{table}
\FloatBarrier
\subsection{Supplementary Note 3: NV Relaxometry Experiments with Different Organic Solvents}
\begin{figure}
    \centering
    \includegraphics[scale=0.45]{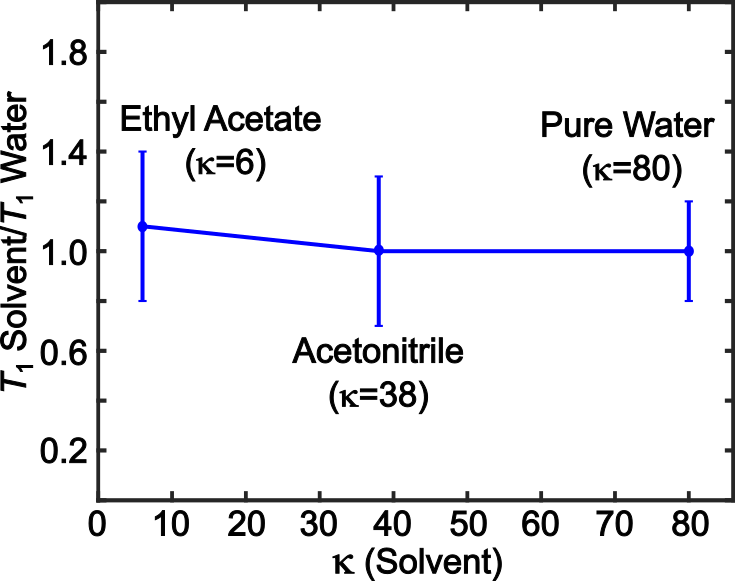}
    \caption{\textbf{NV relaxometry experiments showing the impact of the solvent's dielectric constant ($\boldmath{\kappa}$). Experiments are performed at \textit{f}\textsubscript{\text{NV}} = 1.88 GHz.}}
    \label{fig:FigureS3}
\end{figure}

Following measurements are performed in order to investigate the impact of the solvent's physical properties on NV relaxometry experiments. Therefore, we choose organic solvents with dielectric constants ($\kappa$) which differ significantly from the properties of water.\cite{Seyferth1987} The diamond is covered three times alternatingly with water and the organic solvent. $T_1$ times of the solvents are then normalized to the $T_1$ time of water. Figure \ref{fig:FigureS3} shows that the $T_1$ time remains unaffected by the solvent.
\FloatBarrier
\subsection{Supplementary Note 4: NV Relaxometry Measurement Series of Para- and Diamagnetic Electrolyte Solutions in Increasing Concentrations}
\begin{figure}
    \centering
    \includegraphics[scale=0.40]{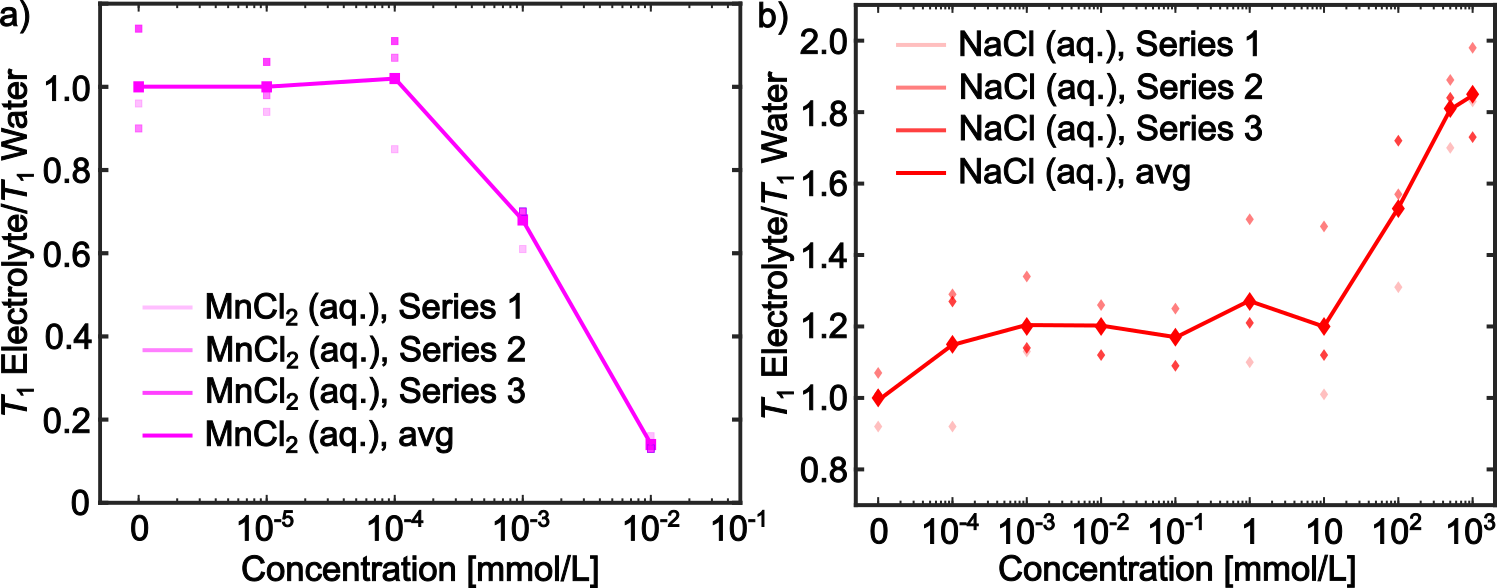}
    \caption{\textbf{NV relaxometry measurement series with increasing concentrations of a) MnCl\textsubscript{2} and b) NaCl solutions. Data points are $T_1$ times normalized to the $T_1$ time of water for each series. Solid lines connect the mean values of three consecutive performed series. Experiments are performed at \textit{f}\textsubscript{\text{NV}} = 1.88 GHz.}}
    \label{fig:FigureS4}
\end{figure}
The NV relaxometry measurement series with paramagnetic MnCl\textsubscript{2} and diamagnetic NaCl solutions are performed in order to determine the sensitivity of the protocol to increasing electrolyte solutions in each case. Paramagnetic MnCl\textsubscript{2} solutions decrease the $T_1$ time in $\sim$ nano- to micromolar concentrations with respect to water. In contrast to that, diamagnetic NaCl solutions increase the $T_1$ time in $\sim$ millimolar concentrations.

\begin{figure}
    \centering
    \includegraphics[scale=0.45]{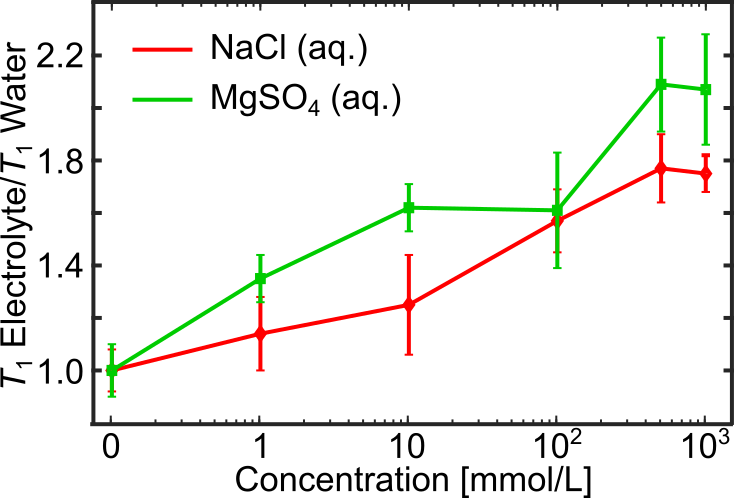}
    \caption{\textbf{NV relaxometry measurement series with increasing concentrations of NaCl (red) and MgSO\textsubscript{4} (green) solutions. Data points are $T_1$ times normalized to the $T_1$ time of water for each series. Experiments are performed at \textit{f}\textsubscript{\text{NV}} = 1.88 GHz.}}
    \label{fig:FigureS5}
\end{figure}
Additionally, we perform NV relaxometry series with monovalent NaCl and divalent MgSO\textsubscript{4} solutions to uncover the influence of differently charged ions (see Figure \ref{fig:FigureS5}). While the data suggests that MgSO\textsubscript{4} has a more significant impact on increasing the $T_1$ time than NaCl, more data and simulations will be needed to draw a definitive conclusion regarding the effect. 

Experiments herein are conducted using the SQ relaxometry pulse sequence (see Methods for detail). We perform each series three times resulting in a mean value for each concentration.
\FloatBarrier
\subsection{Supplementary Note 5: NV Relaxometry Experiments with Differently Oxidized Diamond Surfaces}
\begin{figure}
    \centering
    \includegraphics[scale=0.38]{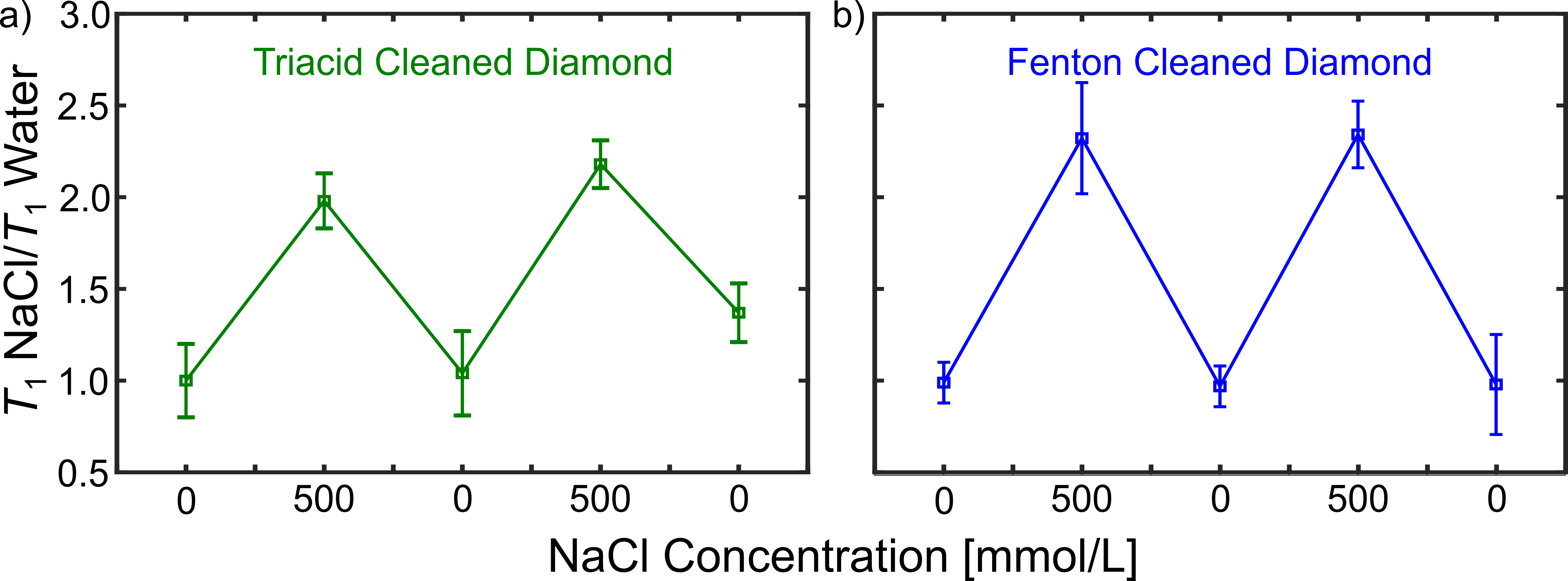}
    \caption{\textbf{NV relaxometry experiments with water and NaCl (500 mM) solutions covering the diamond surface. Diamonds are prepared with the a) triacid clean or b) Fenton reagent. Experiments are performed at \textit{f}\textsubscript{\text{NV}} = 1.88 GHz.}}
    \label{fig:FigureS6}
\end{figure}
We perform NV relaxometry measurements with water and NaCl (\SI{500}{\milli\molar}) solution on two oxidized diamond surfaces that are treated with different reagents in order to investigate the effect on the $T_1$ time behavior (see Figure \ref{fig:FigureS6}). One surface is prepared with the triacid clean procedure according to Brown \textit{et al.}\cite{Brown2019} (see also Figure 3a in the main text), the other one using Fenton chemistry.\cite{Martin2009}
Diamonds are covered alternatingly with water and NaCl solution. Both surface treatments lead to similar results.
\FloatBarrier
\clearpage
\subsection{Supplementary Note 6: NV Depth Dependence Measurements with Water/LiCl (500 mM)}
\begin{figure} [htbp]
    \centering
    \includegraphics[scale=0.38]{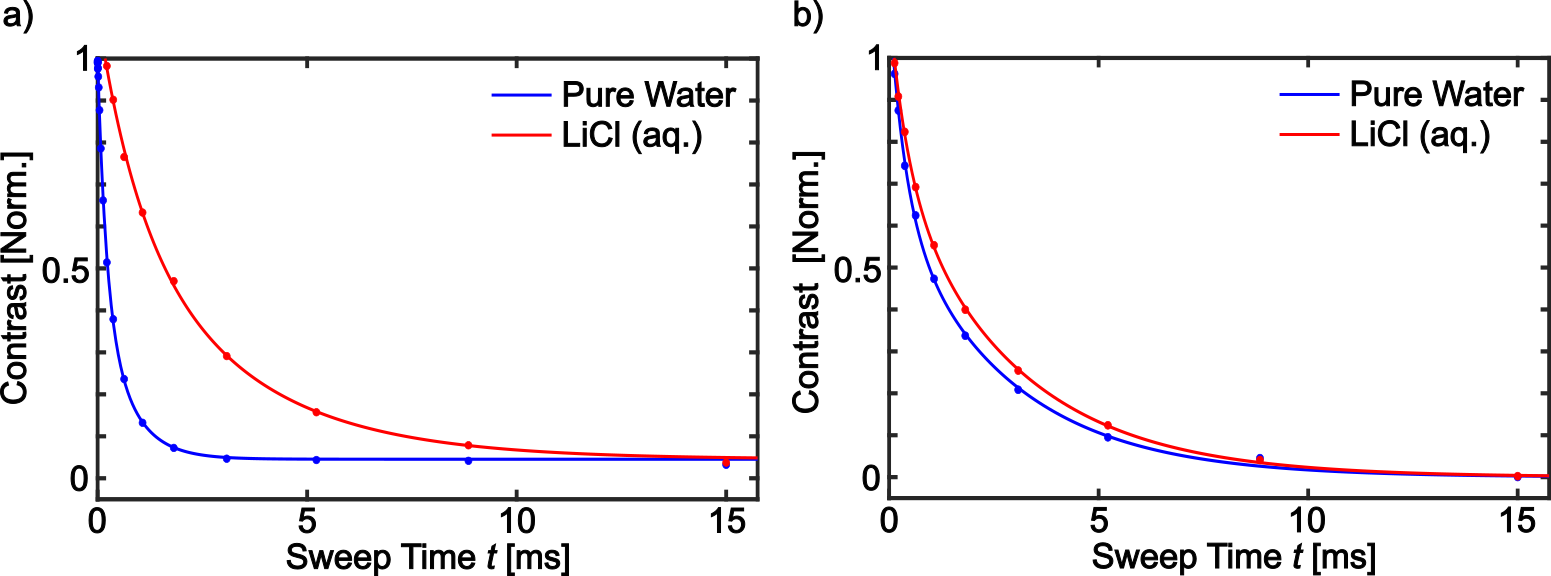}
    \caption{\textbf{$T_1$ relaxation curves of water and LiCl (500 mM) solutions covering the diamond surface. Diamonds were implanted with \textsuperscript{15}N at an energy of a) 2.5\,keV and b) 4\,keV. Experiments are performed at \textit{f}\textsubscript{\text{NV}} = 1.88 GHz.}}
    \label{fig:FigureS7}
\end{figure}
NV relaxometry with water/LiCl  (\SI{500}{\milli\molar}) covering the diamond is performed in order to investigate the impact of another diamagnetic electrolyte and to support the experiments with NaCl (\SI{500}{\milli\molar}) solution using differently deep NV center ensembles (implanted with 2.5\,keV and 4\,keV, see Figure 3b in the main text). Figure \ref{fig:FigureS7} and Table \ref{tab:TableS3} show similar results for both NaCl and LiCl (\SI{500}{\milli\molar}) solution.
 \begin{table}[htbp]
    \caption{\textbf{$T_1$ time constants ($T_{1a}$ and $T_{1b}$) of water, NaCl (500 mM) and LiCl (500 mM) solution on the diamond surface depending on the nitrogen implantation energy. Experiments are performed at \textit{f}\textsubscript{\text{NV}} = 1.88 GHz.}}
    \begin{tabular}{llll}
     \hline
    Implantation energy [keV]&&$T_{1a}$ [\si{\micro\second}]&$T_{1b}$ [\si{\micro\second}]\\
      \hline
2.5&Water&$940\pm180$&$130\pm20.0$\\
&NaCl \SI{500}{\milli\molar}&$1920\pm200$&$170\pm20.0$\\
&Water&$660\pm180$&$200\pm50.0$\\
&LiCl \SI{500}{\milli\molar}&$2940\pm720$&$930\pm40.0$\\
 \hline
4&Water&$1090\pm190$&$190\pm30.0$\\
&NaCl \SI{500}{\milli\molar}&$1270\pm180$&$220\pm40.0$\\
&Water&$2750\pm340$&$380\pm80.0$\\
&LiCl \SI{500}{\milli\molar}&$2880\pm270$&$440\pm90.0$\\
     \hline
    \end{tabular}
    \label{tab:TableS3}
\end{table}
\FloatBarrier
\subsection{Supplementary Note 7: NV Charge State, Coherence and Dephasing Measurements}
\begin{figure} [htbp]
    \centering
    \includegraphics[scale=0.38]{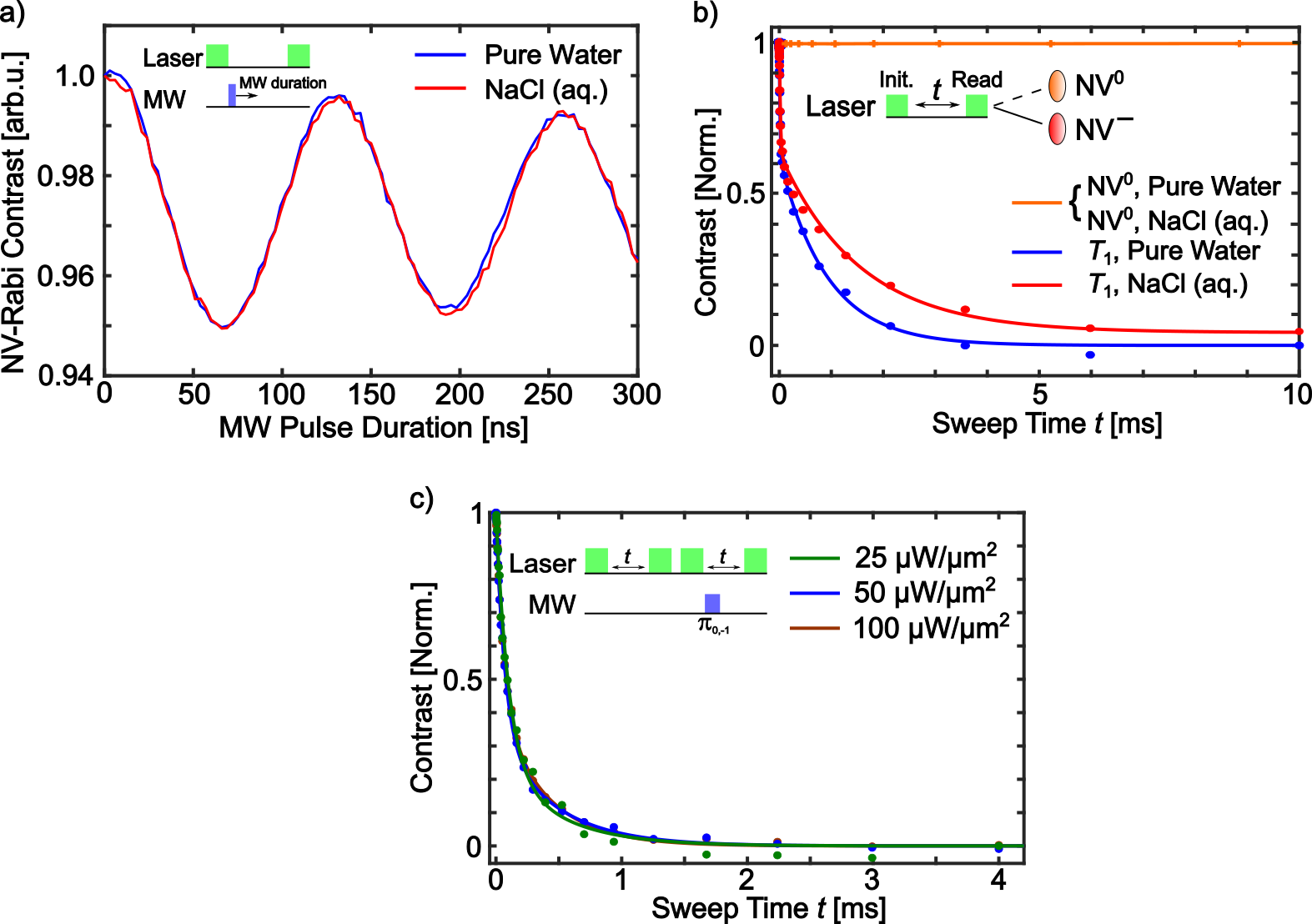}
    \caption{\textbf{Pulse sequences and spectra of NV charge state measurements. a) NV-Rabi experiments, b) NV charge state measurements with selective readout of the NV\textsuperscript{0} or the NV\textsuperscript{--} state and c) $T_1$ relaxation curves using three different laser powers.}}
    \label{fig:FigureS8}
\end{figure}
We observe an increase of $T_1$ by diamagnetic electrolyte solutions. However, it is known that NV charge state alteration (\textit {i.e.}, NV\textsuperscript{0} $\leftrightarrow$ NV\textsuperscript{--}) can influence the outcome of NV relaxometry measurements.\cite{Bluvstein2019a,Dhomkar2018} For that reason, we perform NV-Rabi experiments with water/NaCl (\SI{500}{\milli\molar}) solution (see Figure \ref{fig:FigureS8}a). Any change in the NV-Rabi contrast indicates an alteration of the NV center's charge state. For instance, an ionization of NV\textsuperscript{--} would increase the proportion of NV\textsuperscript{0}, thereby raising the background fluorescence and lowering the contrast. The NV-Rabi experiments show no difference in the outcome between water and the electrolyte implying a constant charge state distribution during the measurement. Secondly, to supplement the NV-Rabi experiments, we conduct NV relaxometry with distinct optical readout of the NV\textsuperscript{0} and NV\textsuperscript{--} charge states and with three different laser powers (see Figure \ref{fig:FigureS8}b and Figure \ref{fig:FigureS8}c). Possible ionization of NV\textsuperscript{--} in the dark or recombination processes would be visible as an alteration in the readout signal of the NV\textsuperscript{0} charge state (see Figure \ref{fig:FigureS8}b).\cite{Bluvstein2019a,Dhomkar2018} These measurements are carried out using the first half of the relaxometry pulse sequence (\textit {i.e.}, without a $\pi$-pulse) and with two different optical filters. The \SI{647}{\nano\meter} long pass filter predominantly reads out the fluorescence from the NV\textsuperscript{--} state and the $600\pm 40$ band pass filter mostly reads out the fluorescence from the NV\textsuperscript{0} state.\cite{Doherty2013a} While a $T_1$ fluorescence decay curve can be extracted from the measurements with the long pass filter, no decisive change in the NV\textsuperscript{0} state is visible using the band pass filter. Probable impact of the laser power on the NV\textsuperscript{--}/NV\textsuperscript{0} ratio and a subsequent change in the $T_1$ relaxation curves is probed with relaxometry experiments using laser powers of 25, 50 and 100 \si{\micro\watt\per\square\micro\meter} (see Figure \ref{fig:FigureS8}c). Both NV-Rabi and NV charge state experiments do not show an impact on NV charge state alteration on the relevant timescales of the relaxometry measurements we conduct herein.
\FloatBarrier
\begin{figure} [htbp]
    \centering
    \includegraphics[scale=0.38]{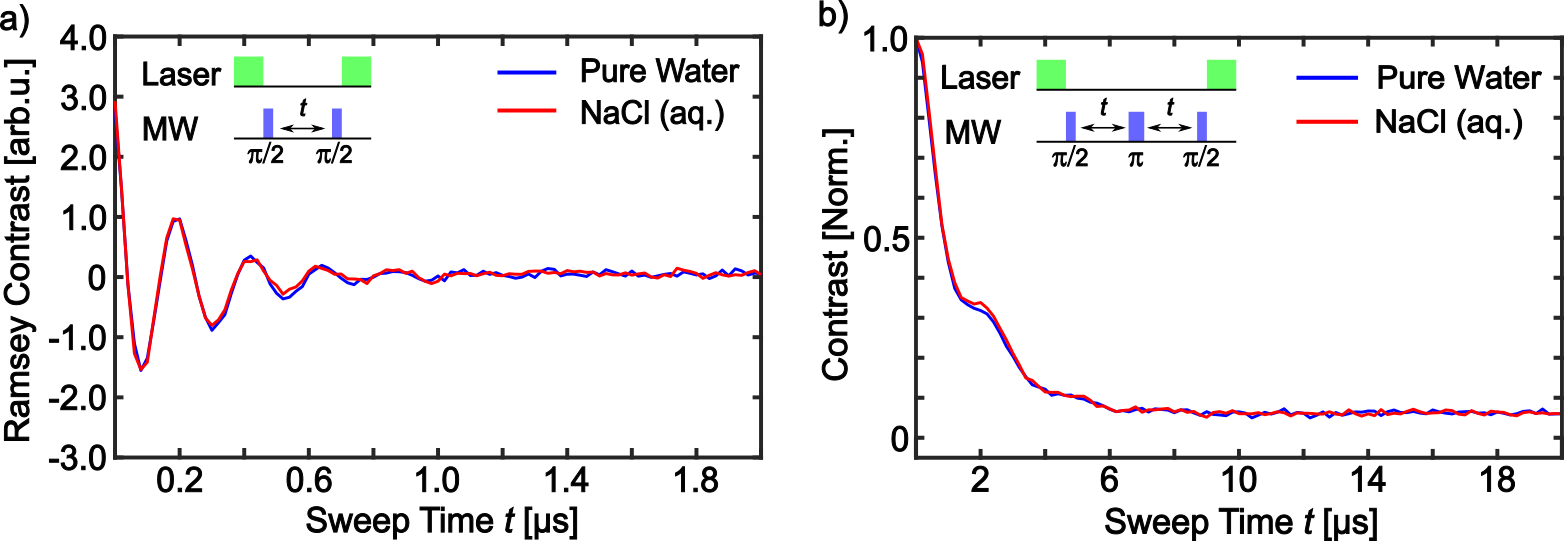}
    \caption{\textbf{Pulse sequences and spectra of Ramsey and $T_2$ Hahn-echo measurements. a) Ramsey experiments performed at a 4 MHz detuned NV\textsubscript{0,-1} resonance frequency $f\textsubscript{\text{NV}}$ and b) $T_2$ Hahn-echo experiments with water and NaCl (500 mM) solution covering the diamond surface at \textit{f}\textsubscript{\text{NV}} = 1.88 GHz.}}
    \label{fig:FigureS9_new}
\end{figure}
Additionally, we perform Ramsey (NV dephasing) and $T_2$ (NV coherence) Hahn-echo experiments, whose outcome is typically affected by changes in the low frequency components of the noise (see Figure \ref{fig:FigureS9_new}a and \ref{fig:FigureS9_new}b).\cite{Rosskopf2014a} Both experiments show no difference in the outcome for water or NaCl (\SI{500}{\milli\molar}) solution. However, we note that probable changes in this noise frequency regime might not be observable with the high-dense NV center ensemble we use in this work, since the surrounding spin-bath (\textit {e.g.}, P1-centers or other paramagnetic impurities) is limiting the NV dephasing and NV coherence in this case.\cite{Barry2020,JacobHenshaw}
\FloatBarrier
\subsection{Supplementary Note 8: \boldmath{$T_1$} Time Constants for Single Quantum and Double Quantum Experiments at 15 G and Zero Field ESR Measurements}
\begin{figure}
    \centering
    \includegraphics[scale=0.4]{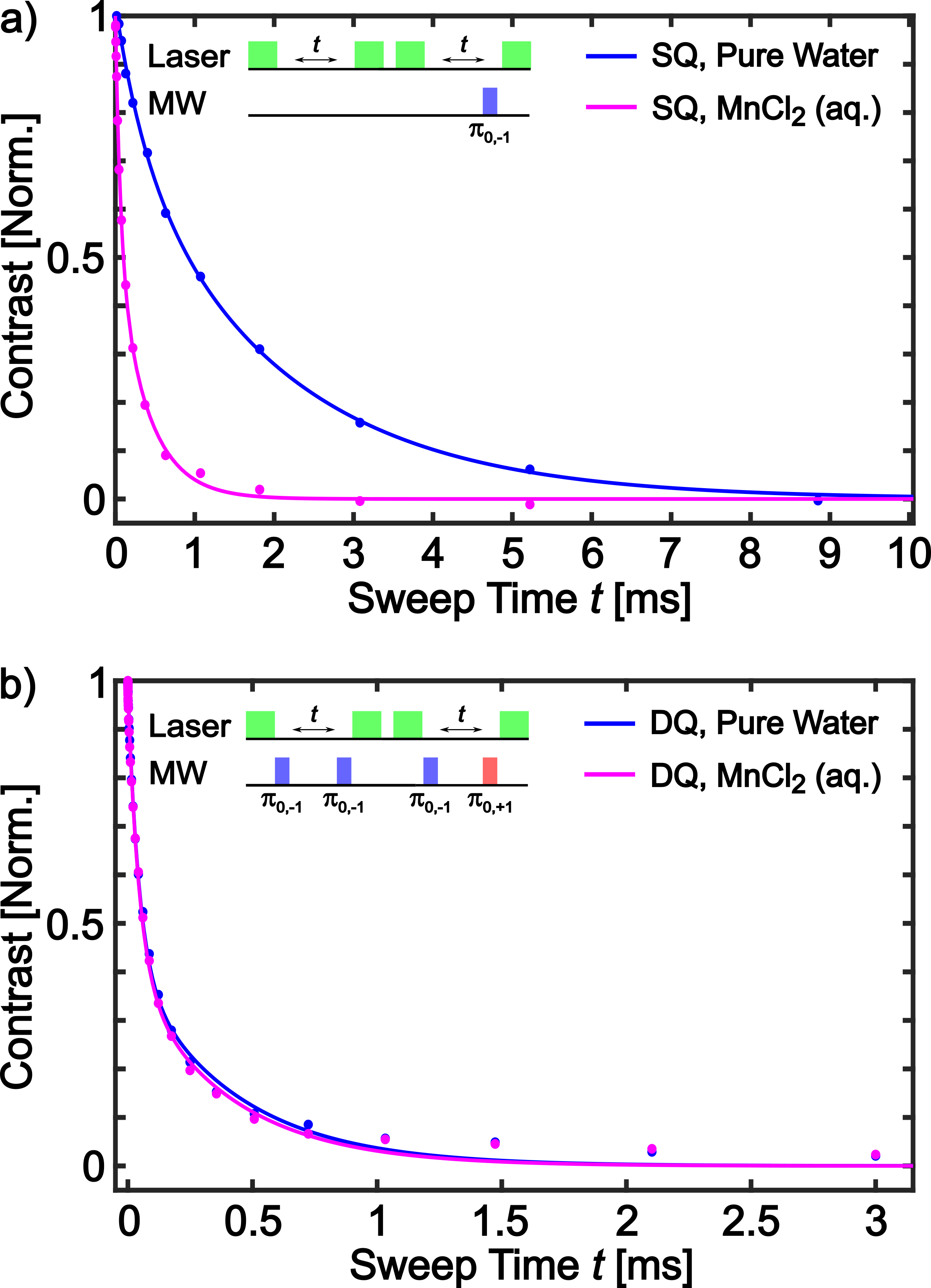}
    \caption{\textbf{a) SQ and b) DQ relaxation curves of water and MnCl\textsubscript{2} (100 \si[detect-weight]{\micro\molar}) solution covering the diamond. Experiments are performed at $B_0$ = 15 G, where the NV\textsubscript{0,-1} transition is at \textit{f}\textsubscript{\text{NV}} = 2.83 GHz (corresponding to a DQ transition frequency of 80 MHz). $T_{1,\text{SQ}}$ decreases by 80\%, whereas $T_{1,\text{DQ}}$ remains unchanged compared to water when MnCl\textsubscript{2} solution covers the diamond.}}
    \label{fig:FigureS10}
\end{figure}

 \begin{table}[htbp]
    \caption{\textbf{$T_1$ time constants (\textit{T}\textsubscript{1a,SQ} and \textit{T}\textsubscript{1a,DQ}) for water/NaCl (500 mM) and water/MnCl\textsubscript{2} (100 \si[detect-weight]{\micro\molar}) solution covering the diamond surface. Experiments are performed at $B_0$ = 15 G.}}
    {\begin{tabular}{lll}
     \hline
    &\textit{T}\textsubscript{1a,SQ} [\si{\micro\second}]&\textit{T}\textsubscript{1a,DQ} [\si{\micro\second}]\\
    \hline
Water&$2600\pm280$&$440\pm24.0$\\
NaCl \SI{500}{\milli\molar}&$3970\pm400$&$1250\pm120$\\
\hline
    Water&$2000\pm340$&$410\pm71$\\
    MnCl\textsubscript{2} \SI{100}{\micro\molar}&$390\pm71.0$&$390\pm78.0$\\
    \hline
    \end{tabular}}
    \label{tab:TableS6}
\end{table}
Single and double quantum $T_1$ experiments of water/NaCl (\SI{500}{\milli\molar}) and water/MnCl\textsubscript{2} (\SI{100}{\micro\molar}) solution covering the diamond surface are performed in order to elucidate the effect of the electrolyte on magnetic and electric field noise. In the case of the NaCl solution, both $T_1$ time constants (\textit{T}\textsubscript{1a,SQ} and \textit{T}\textsubscript{1a,DQ}) increase compared to water, indicating a reduction of both magnetic and electric field noise (see also Table \ref{tab:TableS6}). Importantly, MnCl\textsubscript{2} only reduces the $T_1$ time for the SQ relaxation, whereas the DQ transition remains unaffected compared to water (see also Table \ref{tab:TableS6}). This indicates an exclusive impact of the paramagnetic electrolyte on magnetic field noise. However, we note that probing MnCl\textsubscript{2} in higher ($> \SI{100}{\micro\molar}$) concentrations would lead to a collapse of the NV center's $T_1$ time (see also Figure 2 in the main text). Therefore, a final statement on the impact of higher concentrated paramagnetic electrolyte solutions on the DQ (as well as the SQ) relaxation cannot be made.
\FloatBarrier

\begin{figure} [htbp]
    \centering
    \includegraphics[scale=0.40]{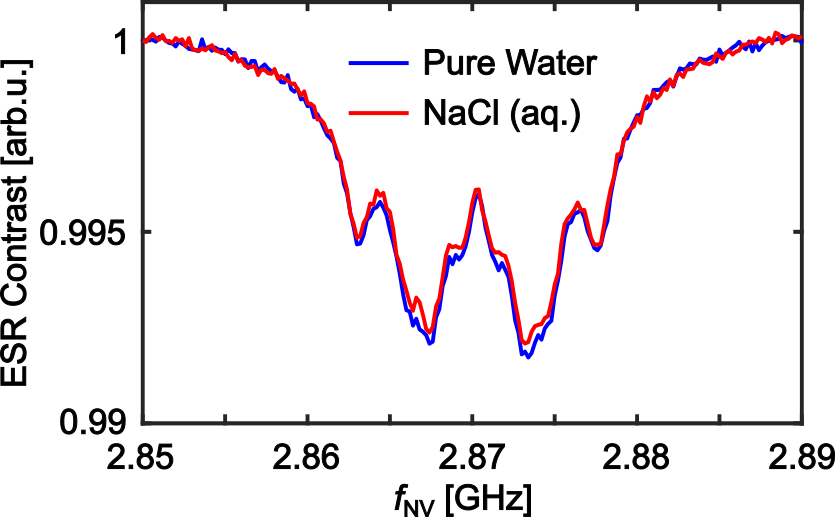}
    \caption{\textbf{ESR experiments at zero magnetic field with water and NaCl (500 mM)  solution covering the diamond surface.}}
    \label{fig:FigureS11}
\end{figure}

\begin{figure}
    \centering
    \includegraphics[scale=0.65]{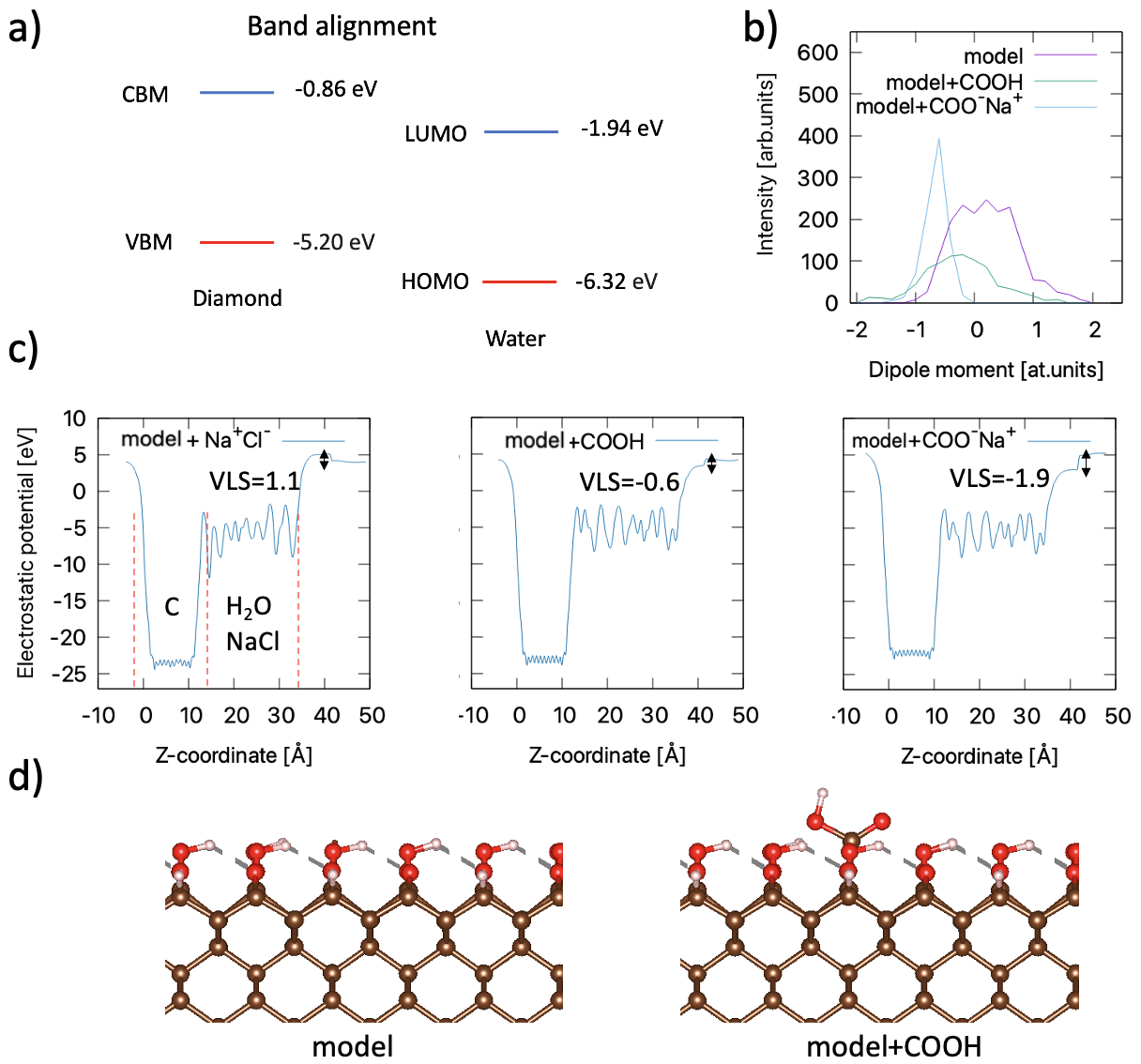}
    \caption{\textbf{a) Band alignment of the water layer and the model diamond surface. b) Distribution of interfacial dipoles, sampled from the MD trajectories for three different compositions of the interface and solvent. c) Average electrostatic potentials and vacuum level shifts (VLS) computed for the configurations corresponding to the middle of the distributions in b. Vertical lines show the parts of the simulation box, spanned by diamond (C), water or aqueous NaCl solution and vacuum. d) Structures of the model diamond surface before and after adding a COOH group.}}
    \label{fig:FigureS12}
\end{figure}

Additionally, we investigate the static electric field environment of the NV center, \textit {i.e.}, charges within the diamond and adjacent to the NV center (\textit {e.g.}, N\textsuperscript{+} and NV\textsuperscript{--}).\cite{Mittiga2018} Therefore, we measure ESR at zero magnetic field (here the earth's magnetic field $\sim$ \SI{0.5}{\Gauss}), because any difference in the static electric field in the proximity of the NV center with respect to water or the electrolyte solution covering the surface would induce a shifting and/or splitting of the $m_{s}=\pm 1$ states apparent in the ESR spectra.\cite{Mittiga2018} Figure \ref{fig:FigureS11} shows no significant change of the ESR resonance lines for the exposure of water or electrolyte solution, indicating that static electric fields do not contribute.
\FloatBarrier
\clearpage
\subsection{Supplementary Note 9: Results of DFT-PBE \textit{Ab Initio} Molecular Dynamics Simulations}
\FloatBarrier

\bibliography{citations}
\end{document}